\documentclass[twocolumn,showpacs,floatfix]{revtex4-1} 
 
\pdfoutput=1
\usepackage{graphicx} 
\usepackage{dcolumn} 
\usepackage{amsmath} 
\usepackage{epsfig} 
\usepackage{amssymb} 
\usepackage{graphics} 
\usepackage{subfig}

\newcommand{\abs}[1]{\left\vert#1\right\vert} 
 
\newcommand{\ket}[1]{\left\vert#1\right\rangle} 
\newcommand{\bra}[1]{\left\langle#1\right\vert}

\newcommand{\beq}{\begin{equation}} 
\newcommand{\eeq}{\end{equation}} 
\newcommand{\bea}{\begin{eqnarray}} 
\newcommand{\eea}{\end{eqnarray}} 
 
\newcommand{\tr}{\mbox{Tr}} 
 
\makeatletter 
\def\btt#1{\texttt{\@backslashchar#1}} 
\DeclareRobustCommand\bblash{\btt{\@backslashchar}} 
\makeatother 
\def\Bid{{\mathchoice {\rm {1\mskip-4.5mu l}} {\rm 
{1\mskip-4.5mu l}} {\rm {1\mskip-3.8mu l}} {\rm {1\mskip-4.3mu l}}}}

\begin{document} 
 
\title{Effects of Noise, Correlations and errors in the preparation of initial states in Quantum Simulations} 
\author{Nayeli Zuniga-Hansen, Yu-Chieh Chi, Mark S. Byrd} 
\email{mbyrd@siu.edu} 
\affiliation{Physics Department and Computer Science Department,  
Southern Illinois University,  
Carbondale, Illinois 62901-4401}

\date{\today} 
 
\begin{abstract} 
In principle a quantum system could be used to simulate another quantum system.   
The purpose of such a simulation would be to obtain information about 
problems which cannot be simulated with a classical computer due to 
the exponential increase of the Hilbert space with the size of the 
system and which cannot be measured or controlled in an actual experiment.  
The system will interact with the surrounding environment, with the other 
particles in the system and be implemented using imperfect controls making it subject 
to noise.  It has been suggested that noise does not need to be controlled to the 
same extent as it must be for general quantum computing.  However the effects of 
noise in quantum simulations and how to treat them are not completely understood.   
In this paper we study an existing quantum algorithm for the one-dimensional 
Fano-Anderson model to be simulated using a liquid-state NMR device. We calculate 
the evolution of different initial states in the original model, and then we add  
interacting spins to simulate a more realistic situation.  
We find that states which are entangled with their environment, and sometimes 
correlated but not necessarily entangled have an evolution which is described 
by maps which are not completely positive.  We discuss the conditions for this to occur and also the implications.  
\end{abstract} 
 
\pacs{03.65.Yz,03.67.Ac}

\maketitle 
 
%\tableofcontents 
 
%------------------------------------------------------------------------- 
%------------------------------------------------------------------------- 
 
\section{Introduction} 
 
Simulating quantum systems with quantum systems is one of the primary
reasons there is a great deal of interest in building a quantum computing device.  
The difficulty of simulating quantum systems on a classical  
computer, mainly due to the exponential increase 
of the Hilbert space with system size, was Richard  
P. Feynman's motivation for proposing the idea that a quantum system 
might perform this task much more efficiently \cite{Feynman:QC}. 
Lloyd showed later that some quantum systems 
could be manipulated to represent the evolution of other quantum 
systems using only local interactions \cite{Lloyd:96}.  
 
There are many problems of interest in quantum mechanics which have no
known analytical solution.  Thus for a wide range of physical systems 
simulation is a valuable tool for solving quantum mechanical problems.   
Classical simulation of such systems can quickly become intractable as
the number of particles increases. 
The resources that are required to perform such a task increase 
exponentially with the size of the system. 
For example, in order to represent the state of $N$ 2-state particles a 
$2^{N}$ vector is required and for its evolution the unitary will be  
a $2^{N}\times2^{N}$ matrix \cite{Lloyd:96,BrownUK:10}.  
However, only $N$ particles would be necessary to simulate such a
system \cite{Lloyd:96,Pritchett:01}.  In this sense, a quantum simulator is 
conjectured to  
provide exponential speedup over classical simulation 
\cite{Boghosian:97b}.   But that is not the only advantage; other
problems such as the sign problem from Quantum 
Monte Carlo algorithms for fermionic systems, or the  
exchange-correlation functionals in Density Functional  
Theory \cite{Ortiz:01,Biamonte:01} will not be present in 
a quantum simulation.  Therefore, many difficult problems in particle
physics, condensed matter systems, and chemistry, among others, could 
be tackled 
\cite{Wiesner,Zalka,Schack:98,Somaroo:99,Kassal1:08,Ortiz:01,Somma:02,Somma:03, 
Negrevergne/etal:05,Simon:11,Clark:09,Barreiro:01,BrownUK:11,Boghosian:97b,Bravyi:08}. 

Quantum simulations have received a great deal of recent attention since 
they are feasible without the need for a universal quantum computing device.  
The question of the universality of Hamiltonians 
has been addressed to a great extent 
\cite{Bennett:01,Dodd:01,Wocjan:01,Wocjan/etal:01,Wocjan:02,Nielsen/etal:02,  
McKague:09,Childs:01,Childs1:10,Childs:11} and 
algorithms have been developed to simulate specific systems \cite{Ortiz:01,Wu/etal:BCS,Pritchett:01, 
Mostame:08,Mostame:11, 
HWang:08,Whitfield:11,BrownUK:11,Yung:10,Kassal1:08,Kassal:09,Lanyon:10,Wang:11,BrownUK/etal:11}.
In addition, experiments have been designed and implemented \cite{Du:10,Simon:11,XSMa:11,JCho:08, 
Roos:11,Gerritsma:10,Gerritsma:11,Johanning:09}. 
However, a great deal of work remains to be done.  
Currently available quantum simulating devices  
have relatively few controllable particles. 
They are, after all, quantum systems that inevitably interact 
with the surrounding environment and therefore are subject to noise.  
Just as with quantum computing, this is an important issue when
it comes to scalability.  It is therefore necessary  
to study how the interactions affect a quantum simulation. 
 
The purpose of the present work is to study effects of noise in  
an existing algorithm proposed for a quantum simulation and to take away from 
this example as much general understanding as we can.   The primary noise 
considered is prior unknown correlations or entanglement  
within the system and between the simulated system and the environment.  
We study the evolution of different initial states, 
including ideal ones and states in which errors are present due to 
mistakes in preparation and/or interactions with particles in  
the system and find the dynamical maps that represent the evolution. 
The algorithm we explore was proposed and developed by 
Ortiz et al. \cite{Ortiz:01} to simulate the one-dimensional Fano-Anderson 
model.  To examine various behaviors of the system with initial
correlations, we first provide a background for the quantum simulation
in Section \ref{sec:introsims} which focuses on the different 
sources of noise that can affect the experiments.  
Section \ref{sec:opensys} provides a brief review of 
open system quantum dynamics, and discusses dynamical maps and  
their main characteristics, including requirements 
for positivity and complete positivity; the purpose is  
to use dynamical maps to describe general errors in simulations.   
Section \ref{sec:alg} contains a brief explanation of the algorithm used,  
including the modifications we made to represent noise in the system. 
Finally, our results, given in Section \ref{sec:concl}, 
are divided in two parts: those states for which the 
Bloch vector only has a component along the $z$ direction, 
and those which have some 
small component along $x$ and a main one along $z$.  We will also discuss 
why this is important.  
These two last subsections are subsequently divided into simulations 
performed with no external noise and simulations with noise. 
For the purpose of comparison, the parameters of the system were obtained  
from Ref.~\cite{Ortiz:01} and were used for all the considered scenarios. 
 
%------------------------------------------------------------------------- 
 
\subsection{Quantum simulations} 
 \label{sec:introsims} 

There are two classifications of quantum simulators.   
The Universal Quantum Simulator (UQS) \cite{Kassal:01} (also referred to  
as Digital \cite{Buluta:01}), 
is a quantum computer represented by the standard circuit model with the 
set of universal gates that act on a collection of two-state systems
\cite{Deutsch:85,Dodd:01,Pavlov:06}.  
The term \textit{universal} implies that the quantum computer  
would be able to simulate any arbitrary quantum  
system \cite{Tseng:00} which implies universal quantum computation is 
possible.   However, a fully functioning 
quantum computer has not been built yet.  
So researchers have designed and implemented  
devices consisting of smaller and controllable quantum systems specifically 
intended for simulations.  This is the other type of quantum simulators,
referred to as Specialized Quantum Simulators (SQS) \cite{Kassal:01,Pravia:01} 
or analogue quantum simulators \cite{Buluta:01,Kendon:01} since they
are not able to be used to simulate any quantum device or
computation.  Rather, they are able to simulate a smaller, but
interesting class of physical systems.   
Examples of such systems include: ultracold atoms, ion traps, 
quantum dots, atoms in optical lattices, coupled cavities, photons, 
 electrons floating on He films and NMR devices    
\cite{Buluta:01,CoryNMR:00,Zhang/etal:05,Pritchett:01,Simon:11,Barreiro:01,BrownUK:11, 
XSMa:11,Johanning:09,JCho:08,Mostame:08}.  
In the SQS, universality for all quantum systems is not required, thus
many interesting advances and simple simulations have already been
performed 
\cite{Jane:03,Zhang/etal:05,Cappellaro/etal:07, 
Vollbrecht:09,Barreiro:01, Du:10,Simon:11,XSMa:11,Roos:11,Gerritsma:10,Lanyon:10,Gerritsma:11}.  
 
Just as is the case with any other quantum system, unwanted
interactions with an environment can have a detrimental effect on the
outcome. Error correction and/or prevention is usually required for accurate implementation.  However, inaccurate unitary transformations are also a source of noise and the evolution of the system under 
a specific Hamiltonian is the main problem  
of interest \cite{Abrams:97,BrownUK:10}. 
 
All steps, preparation, evolution and measurement, can cause some
error \cite{Ortiz:01,Clark:09} as well as 
unwanted interactions with other particles in the simulator, etc. 
It was initially suggested that decoherence in quantum 
simulations may not need to be treated in the same strict sense as in  
quantum computation \cite{Lloyd:96} since noise in the simulating
system might be able to be identified with noise in the simulated
system.  The 
nature of the interactions of the simulator with the bath may not be the  
same as those of the system of interest and thus 
error prevention techniques of some sort will almost certainly be
required.  These 
include error-correcting codes (QECC) \cite{Shor:95,Steane,Calderbank:96, 
Gottesman:97,Gottesman:97b,Gaitan:book},  
decoherence free subspaces/noiseless subsystems (DNS) 
\cite{Zanardi:97c,Lidar:PRL98,Duan:98,Knill:99a,Kempe:00,Lidar:00a}  
(see also \cite{Lidar/Whaley:03,Byrd/etal:pqe04} for reviews),  
and/or dynamical decoupling (DD) 
\cite{Viola:98,Ban,Duan:98e,Viola:99,Zanardi:98b,Vitali:99,Viola:99a,Viola:00a,Vitali:01, 
Byrd/Lidar:01,CoryNMR:00,Agarwal:01}.  However, error correction means an increase in resource requirements, and this can represent a problem with scalability \cite{Kendon:01,Brown:06,BrownUK:10, 
Pritchett:01,Papageorgiou:10} as well as efficiency.  
In addition, there exist algorithms and observables which have an
inherent robustness to errors \cite{Wu/Byrd:algs}, but this is not the case 
for all systems and all errors.

It is sometimes, in fact, possible for the errors to be treated quite differently in the 
simulation of quantum systems.  For example, 
sometimes it is possible to model some of the interactions of an open 
quantum system, as is done in Refs.~\cite{Wang:11,Mostame:11}, 
in which the bath is simulated as well. 
Also, D\"ur et al. propose an algorithm to generate many body   
interactions \cite{Dur:08} to study the influence of noise in the 
simulation of many body systems.  
Furthermore, an experimental setup to study open systems is proposed in \cite{Barreiro:01}. 
In these cases, the environment is included, 
but external interactions will still be present and will affect the 
final outcome in an undesirable manner.
 
Our work examines interactions within the system, where 
errors in preparation of the initial state 
give rise to errors.  
Obviously the initial state is important because the
outcome of the simulation depends on it.  Also, when errors 
are caused by initial entanglement, dynamical decoupling 
cannot remove those errors since these
controls rely on local unitary transformations to
eliminate Hamiltonian interactions with a bath.  Local unitary
controls cannot change the entanglement between the system 
and the bath.  

Experimentally, it has been observed 
that two different state preparation methods may not yield 
the same result and can have a profound effect on the outcome  
\cite{Kuah:07}.  
We observe the characteristics of the dynamical map, 
(which will be described more in detail in the next section) 
that describe the evolution of different initial states 
and determine their positivity or complete positivity.  
Until recently, discussions of the evolution of an 
open quantum system were limited to 
completely positive maps.  However, 
work by Pechukas \cite{Pechukas:94} and more recently by Shaji and Sudarshan  
\cite{Shaji/Sudarshan:05} have provided demonstrations 
that a map does not need be completely positive for the end result 
to represent a physical state. It fact, the map does not even need to
be positive; it must only be positive on a given domain in order to
possibly represent a physical mapping.  
In certain circumstances dynamical maps can provide information about 
correlations in the initial state of the 
system, which could provide useful information about the 
effects of noise and interactions in quantum simulations.
Furthermore, there are many sets of operators in the operator-sum
decomposition which give rise to the same map.  This is true of
completely positive maps \cite{Nielsen/etal:97,Nielsen/Chuang:book}
as well as maps which are not completely positive \cite{Ou/Byrd:10a}. 
 
%------------------------------------------------------------------------- 
 
\subsection{Noise in Quantum Systems, Completely and Non Completely Positive Maps} 
 
\label{sec:opensys} 
 
The density matrix, or density operator, represents our knowledge of 
the quantum state of a system.   
In general any density operator must satisfy the following conditions in order to 
represent a physical state \cite{Sudarshan:61}:   
\begin{equation} 
\label{eq:rho1} 
\rho = \rho ^{\dagger}, \text{ it is Hermitian,}  
\end{equation} 
\begin{eqnarray} 
\rho \geq 0,&& \text{ it is positive semi-definite,} \nonumber \\
&&{\text{ i.e. its eigenvalues are non-negative,}}  \label{eq:rho2} 
\end{eqnarray} 
\begin{eqnarray} 
\text{Tr}(\rho)=1,&& \text{ it has trace $1$,}\nonumber \\
 &&\text{ i.e. the sum of the probabilities is $1$.} \label{eq:rho3} 
\end{eqnarray} 
The evolution of a closed system is described by a unitary transformation, as 
\beq 
\psi(t) = U \psi (0), \nonumber 
\eeq 
where $U=\exp{iHt}$.  It follows that  
\beq 
\rho (t)=U \rho(0) U^{\dagger}. \nonumber 
\eeq 

The density operator is often written as an expansion of pure states  
\beq 
\rho=\sum_j p_j \ket{j}\bra{j},	\nonumber 
\eeq 
where the $p_j$ are the probabilities associated to each of the states 
$\ket{j}$.  If one of the 
probabilities is equal to $1$ and the rest are $0$, then the state is pure.   
For two-state systems we can write the density operator in terms of the  
$2 \times 2$ unit matrix and the Pauli operators,  
\beq 
\rho=\frac{1}{2} \left( \Bid + \vec{a} \cdot \vec{\sigma} \right), \nonumber 
\eeq 
where the coefficients $a_i$ are the projections along the $x$, $y$ 
and $z$ directions of the so-called Bloch vector.    
This provides a representation of the quantum state 
in the well-known Bloch sphere, which is a geometric  
representation of the states of a qbits in terms of a sphere with 
radius $1$.  (For higher dimensional systems, this is referred to as the  
polarization vector, coherence vector, or generalized block vector.  See 
\cite{Mahler:book,Arvind:97,Englert/Metwally,Jakob:01,Byrd/Khaneja:03,Kimura,Byrd/etal:11} 
and references therein.).   
The magnitude of the Bloch vector  
is constrained by the condition $\sqrt{a_{x}^{2}+a_{y}^{2}+a_{z}^{2}} \le 
1$, and $\mid \vec{a}\mid = 1$  
represents a pure state.  Thus any state on the surface of the Bloch sphere 
is a pure state. A mixed state 
is represented by a vector with $\mid \vec{a}\mid < 1$. With this notation 
it is possible to have a visual representation of the 
quantum states at different times.

A system $S$ that is coupled to an environment $E$ with 
Hilbert spaces $\cal H_S$ and $\cal H_E$, respectively, 
can be considered a larger isolated system whose initial state 
is described by $\rho_{SE}(0)$. The time 
evolution of this system is then given by the joint evolution of the system  
and environment \cite{Shaji/Sudarshan:05}  
\beq 
\rho_{SE}(t)= U\rho_{SE}(0) U^\dagger. \nonumber 
\eeq 
We are often only interested in the evolution of the 
system, $S$. Tracing out the environmental  
degrees of freedom provides us with the 
 reduced dynamics of the system  
\beq 
\rho_S (t)= \text{Tr}_E [\rho_{SE} (t)] = \text{Tr}_E \left[U_{SE}(t) \rho_{SE} (0) 
  U_{SE}^\dagger\right]. \nonumber  
\eeq 
With the reduced dynamics of $S$, we can find the map that transforms 
the initial state $\rho(0)$, into the final state $\rho(t)$.   
To obtain the "dynamical map" it is convenient to write the $N\times N$ density operator $\rho$   
as a $N^2\times 1$ column vector that is transformed into another $N^2\times 1$ column vector 
through the  $N^2\times N^2$ supermatrix $A$ 
\begin{equation} 
\label{eq:mapAin} 
\rho_{r^{'}s^{'}}(t)  = A_{r^{'}s^{'},rs} \rho_{rs}(0), 
\end{equation} 
where $A$ describes the most 
general evolution of $\rho$ \cite{Cesar/Sudarshan:08}.  In  
matrix notation 
\begin{equation} 
\label{eq:mapA} 
\rho^{'}= A \rho. 
\end{equation}

Because $\rho$ must be mapped to another positive 
$\rho^{'}$ the following conditions are imposed on $A$ \cite{Sudarshan:61}: 
\begin{equation} 
\label{eq:cond1A} 
A_{r^{'}s^{'},rs}=(A_{s^{'}r^{'},sr})^{*} \textrm{ , is Hermitian,} 
\end{equation} 
\begin{equation} 
\label{eq:cond2A} 
\sum_{rsr^{'}s^{'}}x^{*}_{r}x_{s}A_{rs,r^{'}s^{'}}y^{*}_{r^{'}}y_{s^{'}} 
\geq 0 \textrm{ ,$A$ is positive,}  
\end{equation} 
\begin{equation} 
\label{eq:cond3A} 
\sum_{r}A_{rr,r^{'}s^{'}}=\delta _{r^{'}s^{'}} \textrm{ ,$A$ is Trace Preserving.} 
\end{equation} 
These conditions ensure the conditions
Eqs.~(\ref{eq:rho1})-(\ref{eq:rho3}) on the density operator are satisfied.

By interchanging indices of $A$, we obtain another $N^2\times N^2$ 
supermatrix $B$ \cite{Sudarshan:61}  
\begin{equation} 
B_{rr^{'},ss^{'}} \equiv A_{rs,r^{'}s^{'}}. 
\end{equation} 
The $1 \times N^2$ rows of $A$ become the $N\times N$ 
block matrices of $B$.   
The following conditions are imposed on $B$ so that it represents a 
physical map:  
\begin{equation} 
\label{eq:cond1B} 
B_{rr^{'},ss^{'}}=(B_{r^{'}r,s^{'}s})^{*},\; B \textrm{ is Hermitian,} 
\end{equation} 
\begin{equation} 
\label{eq:cond2B} 
\sum_{rsr^{'}s^{'}}x^{*}_{r}y_{r^{'}}B_{rr^{'},ss^{'}}x_{s}y^{*}_{s^{'}} 
\geq 0, \textrm{ $B$ is positive semi-definite,} 
\end{equation} 
\begin{equation} 
\label{eq:cond3B} 
\sum_{r}B_{rr^{'},rs^{'}}=\delta _{r^{'}s^{'}}, \textrm{ $B$ is trace preserving.} 
\end{equation} 
From these we may write  
\beq  
\label{eq:dmap} 
\rho(t)= B\left[\rho(0)\right].   
\eeq 
If $B$ is decomposed into its eigenvectors and eigenvalues, the action of the map  
can be represented as follows  
\beq 
B\left[\rho(0)\right]=\sum_{\alpha} \lambda_{\alpha} \zeta_{\alpha}\rho(0) \zeta_{\alpha}^\dagger, \nonumber 
\eeq 
where $\lambda_{\alpha} \in \mathbb{R}$ are the eigenvalues. 
The hermiticity of $\rho^\prime$ is guaranteed by the restriction given in 
Eq.~(\ref{eq:cond1B}) \cite{Cesar/Sudarshan:08}, so that $B$ must be   
Hermitian. The matrix $A$ is required to transform $\rho(0)$ into another Hermitian 
state $\rho(t)$, but $A$ is not necessarily Hermitian itself. 
The complete positivity of the map implies that the final state will be positive. 
The eigenvalues of $B$ must all be positive for it to be guaranteed to
be a completely positive map. 
If $B$ has a negative eigenvalue but still transforms  
any positive $\rho(0)$ into a positive $\rho(t)$, then $B$ is a positive but not 
necessarily completely positive map. 

Non-completely positive maps have been measured using quantum process
tomography (QPT) \cite{O'Brien:04,Bendersky:08} which has 
caused the specifics of QPT to be questioned \cite{Cesar/etal:10}.
But the possibility that a map which is not a completely positive map
can transform a valid quantum state into another valid state has
brought a great deal of interest in  
studying the conditions for complete positivity.  This is in addition
to the interest in it due to the partial transpose as an indicator of 
entanglement \cite{Peres,Horodeckis}.  

In 1994, Pechukas showed that complete positivity constrains   
a system to product states of the form   
$\rho_{SE} = \rho_S \otimes \rho_E$, where $\rho_E$ is a fixed state 
of the bath \cite{Pechukas:94,Pechukas+Alicki:95}   
which excludes correlations and does not represent many physical 
situations.  Alicki in Ref.~\cite{Alicki:95} argued that there is no 
general definition for the reduced quantum dynamics beyond the weak 
coupling regime, therefore, when the system is in an initially correlated 
state with the environment, linear assignment maps have no unique definition 
\cite{Cesar/etal:10}, and linearity would only be preserved 
for states that are invariant under the transformation
\cite{Alicki:95}.  
Pechukas replied in Ref.~\cite{Pechukas+Alicki:95}, and agreed that 
open system reduced dynamics can be non-linear.  
However, Rodriguez-Rosario et al. examine the assignment maps in 
\cite{Cesar/etal:10} and argue against giving up linearity 
by noting that the assignment maps can be linear if the conditions 
of consistency or positivity are relaxed, and favor relaxing the 
positivity condition. 
A quantum system that interacts with the environment before our
prescribed $t=0$ can be 
described by completely positive dynamics if the environment  
does not re-act on the system \cite{Cesar/Sudarshan:08}, i.e. the 
coupling is weak and/or the initial state is in a particular 
form \cite{Pechukas:94}.  

As mentioned above, when the map is completely positive 
the eigenvalues of $B$ in Eq.~(\ref{eq:dmap}) can be taken to all 
be positive.  When they are, Eq.~(\ref{eq:dmap}) can be rewritten as   
\beq  
\label{eq:kmap} 
\rho(t)= B\left[\rho(0)\right]=\sum_{\alpha} \lambda_{\alpha} \zeta_{\alpha}\rho(0) \zeta_{\alpha}^\dagger= 
\sum_{\alpha} C_{\alpha}\rho(0) C_{\alpha}^\dagger,  
\eeq 
where $C_{\alpha} = \sqrt{\lambda_{\alpha}} \zeta_{\alpha}$. 
%, and the 
%operators $C_{\alpha}$ satisfy $\sum_{\alpha} C^{\dagger}_{\alpha} 
%C_{\alpha}=\Bid$.  
Eq.~(\ref{eq:kmap}) is sometimes known as the Kraus 
representation or operator-sum decomposition \cite{Tong:04}, although
it was originally discussed in this context by Sudarshan, Mathews, and
Rau \cite{Sudarshan:61}.  
Jordan, et al.~demonstrated that entanglement in the initial state of 
the system can lead to non-completely positive  
maps that still transform a positive $\rho$ into another positive 
$\rho^{'}$ \cite{Jordan:04}. Rodriguez-Rosario, et al.~found 
that for purely classical correlations, 
the ``quantum discord'' (defined below) 
vanishes, and this is a sufficient condition for completely 
positive reduced dynamics \cite{Cesar/etal:08}. 
Later, Shabani and Lidar demonstrated that the quantum discord  
was also a necessary condition for 
complete positivity \cite{Shabani/Lidar:09a}.  
Quantum discord was introduced by Ollivier and Zurek in 2001, it is  
defined as a 'measure of the  
quantumness of the correlations' \cite{Oll/Zurek:01}, and is 
calculated as follows:  
\bea
\label{eq:QDiscord} 
\delta (S:E) =&& - \text{Tr} \left(\rho_E \log(\rho_E)\right)+
\text{Tr} \left(\rho_{SE} \log(\rho_{SE})\right) \nonumber \\ 
              && - \sum_j \text{Tr} (\Pi^{E}_{j} 
\rho_{SE}) \frac{\Pi^{E}_{j} \rho_{SE} \Pi^{E}_{j} }{\text{Tr} (\Pi^{E}_{j} \rho_{SE})}, 
\eea 
where $H(x)= H(\rho_{x})=-\text{Tr} \left(\rho_x \log(\rho_x)\right)$ is the 
Von Neumann entropy  
and $- \sum_j \text{Tr} (\Pi^{E}_{j} \rho_{SE}) \frac{\Pi^{E}_{j} \rho_{SE} 
  \Pi^{E}_{j} }{\text{Tr} (\Pi^{E}_{j} \rho_{SE})}$ is   
the conditional entropy, defined as the entropy of the system with 
respect to a set of projective measurements performed on the environment.   
Quantum discord provides a measure of the nature of correlations,  it vanishes for 
classical correlations and is maximum when there is entanglement.

%------------------------------------------------------------------------- 
 
\section{Background} 
 \label{sec:alg} 

As mentioned before, the extent to which the noise from the environment can 
be included in a quantum simulation is dependent on both the 
simulating and simulated systems.  Of course it would useful to 
have some previous knowledge of the system-bath interactions. 
However, this is often not the case.  Here we study effects of
unwanted noise in a quantum simulation using an 
algorithm \cite{Ortiz:01} that simulates the one dimensional
Fano-Anderson model. In this case we have a realistic model of the
interaction and use the dynamical maps of the system to describe 
the noisy evolution.  Starting with different initial   
states of the system and bath, we reduce the dynamics to a 
two-particle model system.  The algorithm requires the  
two particles to be initialized in a particular state.  Due to 
interactions with external qbits in the simulating device, these
initial conditions may be imperfect.  In addition, if the particles 
are allowed to interact for some small time before the begining of 
the actual algorithm, the particles could begin in a correlated or 
entalged state.  We 
consider the possibility of errors in the preparation of one of  
the particles in the system as well as the possibility of correlations
between particles. 
We added a visualization of the evolution of the Bloch vector in order
to provide an intuitive picture of the differences in the initial states 
and how they evolve. 
It is useful to note that, regardless of the non-complete positivity of some of 
the maps obtained, the final state is a physical state and the 
system is a realistic physical model with realistic couplings.  The  
significance of these results will be discussed in the conclusions.   
We now describe our methods and results.   
 
%------------------------------------------------------------------------- 
 
\subsection{Quantum Algorithm} 
 
Ortiz, et al. proposed an algorithm for the quantum simulation of the 
one-dimensional Fano-Anderson model \cite{Negrevergne/etal:05}.  
This model consists of an impurity described by an energy $\epsilon$ 
surrounded by a ring of $n$ spinless fermions having energies  $\varepsilon_{k_i}$. 
The fermions interact with the impurity, which is also a spinless fermion,  
through a hopping potential $V$ \cite{Negrevergne/etal:05,Ortiz:01}. 
The diagonalized wave-number representation of the Fano-Anderson Hamiltonian 
is given by \cite{Negrevergne/etal:05,Ortiz:01} 
\beq 
\label{eq:f-amodel} 
H = \sum_{i=0}^{n} \varepsilon_{k_i} c_{k_i}^\dagger c_{k_i} + \epsilon b^\dagger b  
+ V \sum_{i=0}^{n-1} (c_{k_i}^\dagger b + b^\dagger c_{k_i}) \delta_{k_i 0}. 
\eeq 
The system is mapped via Jordan-Wigner transformation to the spin
system to obtain \cite{Ortiz:01} 
\beq 
\label{eq:simH} 
\bar{H} = \frac{\epsilon}{2} \sigma_{z}^{1} + \frac{\varepsilon_{k_0}}{2} \sigma_{z}^{2}  
+ \frac{V}{2} (\sigma_{x}^{1} \sigma_{x}^{2} + \sigma_{y}^{1} \sigma_{y}^{2}). 
\eeq 
Ortiz, et al. consider an NMR device for their simulation as do we, but  
the model is certainly not limited to this type of device.   
The simulator has an NMR drift Hamiltonian of the form \cite{Ortiz:01}  
\bea 
\label{eq:NMRH} 
\!\! H_{d}\!\! &=& \!\!\frac{1}{2}\left(\frac{(\epsilon+\varepsilon_{k_0})}{2} 
-\sqrt{\left(\frac{\epsilon-\varepsilon_{k_0}}{2}\right)^2+V^2}\right)\!
\sigma_{z}^1 \nonumber \\
     && \!\!\! +\frac{1}{2}\left(\frac{(\epsilon+\varepsilon_{k_0})}{2}+\sqrt{\left(\frac{\epsilon-\varepsilon_{k_0}}{2}\right)^2+V^2}\right) \!
 \sigma_{z}^2. 
\eea
The control Hamiltonian for spins in the system is 
\beq   
H_c(t) = \sum_j[\alpha_{x_j} \sigma_x + \alpha_{y_j}\sigma_y]
  \sum_{ij}\alpha_{i,j}\sigma_z^i\sigma_z^j,
\eeq
where the $\alpha$ are controllable.  The last term is considered
controllable because it can be turned on/off with the $x$ and $y$
rotations.  

To obtain the representation of the Hamiltonian in Eq.~(\ref{eq:simH}),  
the following control sequence can be applied to  Eq.~(\ref{eq:NMRH}) \cite{Ortiz:01}  
\bea 
\label{eq:controlseq} 
U = && e^{i\frac{\pi}{4}\sigma_x^2}e^{-i\frac{\pi}{4}\sigma_y^1}e^{-i\frac{\theta}{2}\sigma_z^1\sigma_z^2}e^{i\frac{\pi}{4}\sigma_y^1}e^{i\frac{\pi}{4}\sigma_x^1}
\nonumber \\
  &&\times e^{-i\frac{\pi}{4}\sigma_x^2}e^{-i\frac{\pi}{4}\sigma_y^2}e^{i\frac{\theta}{2}\sigma_z^1\sigma_z^2}e^{-i\frac{\pi}{4}\sigma_x^1}e^{i\frac{\pi}{4}\sigma_y^2}. 
\eea 
The goal is to see if the initial state of the impurity has changed over time and,  
if so, how much.  For this purpose, we use the time
correlation function $C(t) = b(t) b(0)^{\dagger}$, which in 
spin operator representation becomes 
$C(t)=e^{i \bar{H}t}\sigma^{1}_{-}e^{-i \bar{H}t}\sigma^{1}_{+} $ \cite{Ortiz:01}, 
where $\sigma_{+}=\sigma_{x}+i\sigma_{y}$ and 
$\sigma_{-}=\sigma_{x}-i\sigma_{y}$.  
The time correlation function provides information about the overlap of
the initial and final states of the impurity.  

We use the same form of the Hamiltonian in
Eq.~(\ref{eq:simH}) to perform the unitary evolution on different
initial states of the system and perform the same operation
regardless of prior interactions. We then obtain the reduced dynamics of
the state of the impurity site (qbit $1$) and then 
obtain the dynamical map that describes the evolution. We
also calculate the time correlation function for the purpose of
comparing the results of the different situations to those of an ideal
scenario. In this way we observe the effects of the noise and
possible errors in the outcome of the simulation.

%------------------------------------------------------------------------- 
 
\subsection{Simulation with Noise} 
 
To include other qbits in the environment surrounding the system of
interest we modified the control Hamiltonian in two different ways: 
\begin{enumerate} 
\item First, we added two spins and had them interacting via $zz$ 
coupling with the particle that represents the state of the fermion 
site (qbit 2): 
\bea 
\label{eq:controlh2s} 
H_{\text{NMR}} &=&
\frac{1}{2}\left(\frac{(\epsilon+\varepsilon_{k_0})}{2}-\sqrt{\left(\frac{\epsilon-\varepsilon_{k_0})}{2}\right)^2
    + V^2}\right)\sigma_{z}^1 \nonumber \\
 && +\frac{1}{2}\left(\frac{(\epsilon+\varepsilon_{k_0})}{2}+\sqrt{\left(\frac{(\epsilon-\varepsilon_{k_0})}{2}\right)^2 + V^2}\right)\sigma_{z}^2 
\nonumber \\ 
&&+\frac{J_{zz}}{4}\sigma_{z}^{2}\sigma_{z}^{3}+\frac{J_{zz}}{4}\sigma_{z}^{2}\sigma_{z}^{4}+\frac{J_{zz}}{4}\sigma_{z}^{3}\sigma_{z}^{4}. 
\eea 
\item Next, we added an extra particle, which interacts 
in the same fashion ($zz$ coupling) with both particles that represent 
the system of interest: the resonant impurity and the fermion site: 
\bea 
\label{eq:controlh1s} 
H_{{\text \scriptsize NMR}} &=&
\frac{1}{2}\left(\frac{(\epsilon+\varepsilon_{k_0})}{2}-\sqrt{\left(\frac{\epsilon-\varepsilon_{k_0})}{2}\right)^2
    + V^2}\right)\sigma_{z}^1 \nonumber \\
       &&+\frac{1}{2}\left(\frac{(\epsilon+\varepsilon_{k_0})}{2}+\sqrt{\left(\frac{(\epsilon-\varepsilon_{k_0})}{2}\right)^2 + V^2}\right)\sigma_{z}^2 
\nonumber \\ 
&&+\frac{J_{zz}}{4}\sigma_{z}^{1}\sigma_{z}^{3}+\frac{J_{zz}}{4}\sigma_{z}^{2}\sigma_{z}^{3}, 
\eea 
\end{enumerate} 
where $J_{zz}$ represents the $zz$ coupling constant.   
We used the same control sequence from Eq.~(\ref{eq:controlseq}) to obtain Eq.~(\ref{eq:simH}), 
to represent a situation in which the extra qbits are 
environmental and thus are taken to be unknown.

%------------------------------------------------------------------------- 
 
\section{Results} 
 
In this section we describe the results of the simulations for the two  
different modifications to the Hamiltonian as well as different initial states.   

%------------------------------------------------------------------------- 
 
\subsection{States with Bloch vector in the z direction} 
 
We first consider states with only a $z$ component to their Bloch  
vectors.  These form a special class of states  
due to the commutativity of the $zz$ Hamiltonian with these  
initial states.
\begin{center}
\begin{figure}[tpb] 
	\centering 
	\subfloat[t=0]{\includegraphics[scale=0.3]{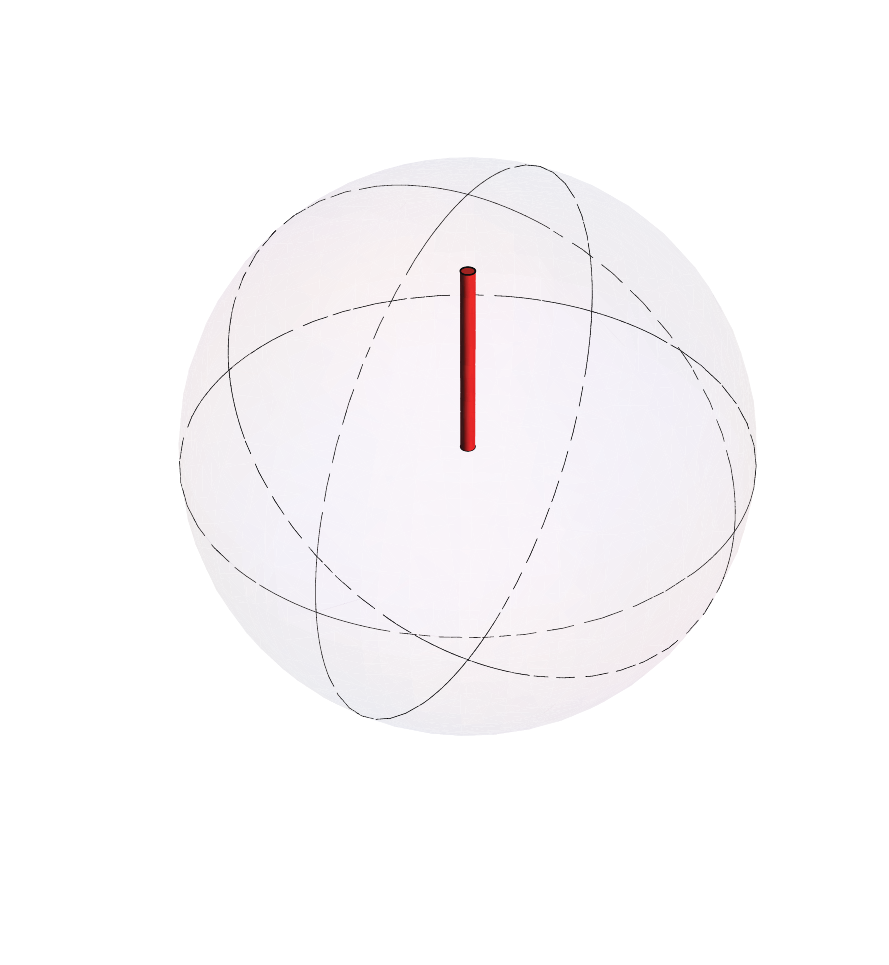}} 
	\subfloat[t=0.3]{\includegraphics[scale=0.3]{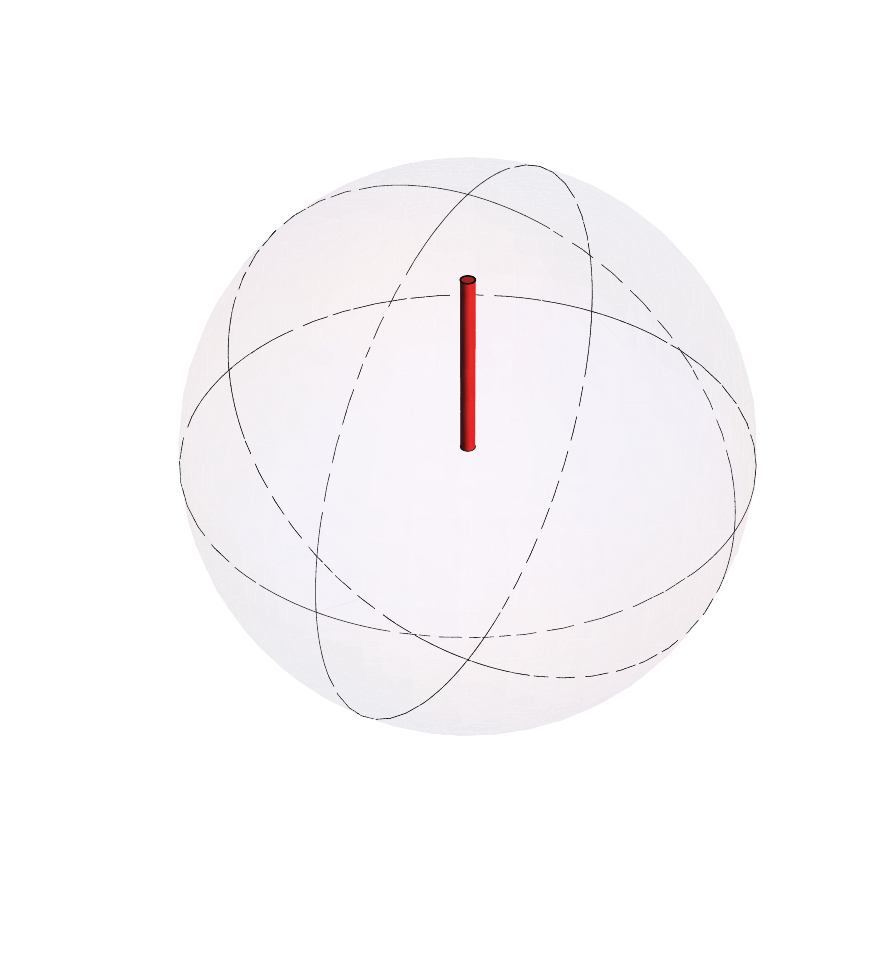}} 
        \quad
	\subfloat[t=0.6]{\includegraphics[scale=0.3]{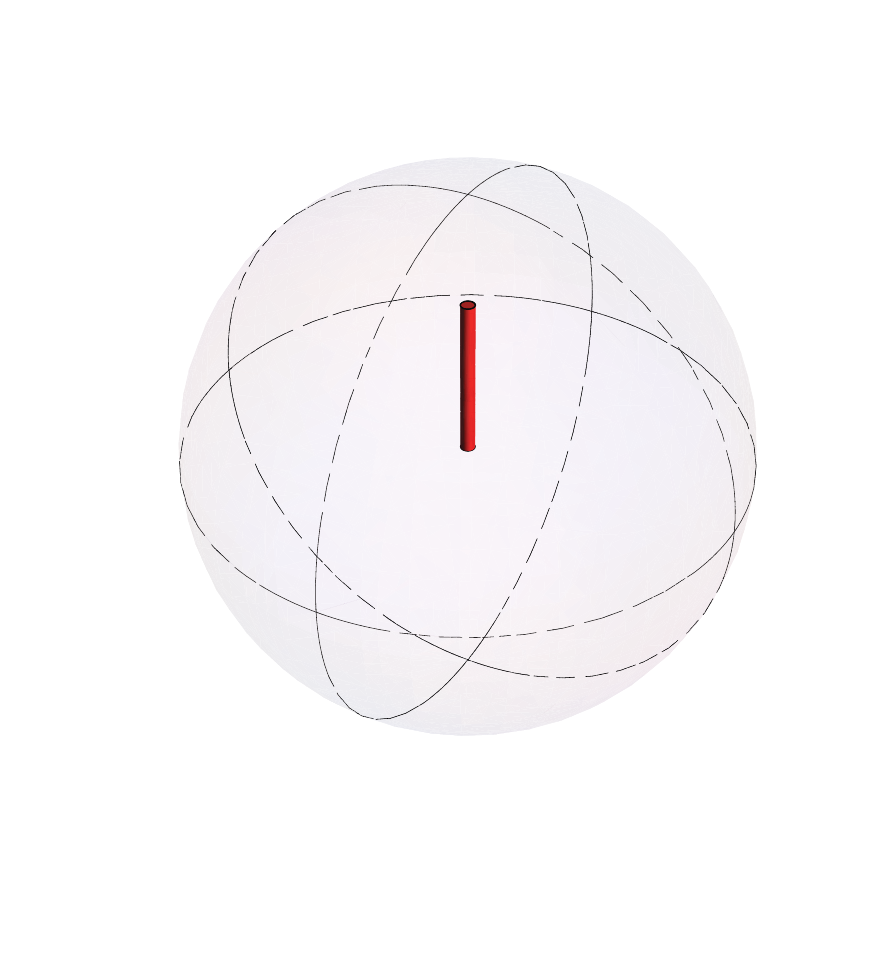}} 
	\subfloat[t=0.9]{\includegraphics[scale=0.3]{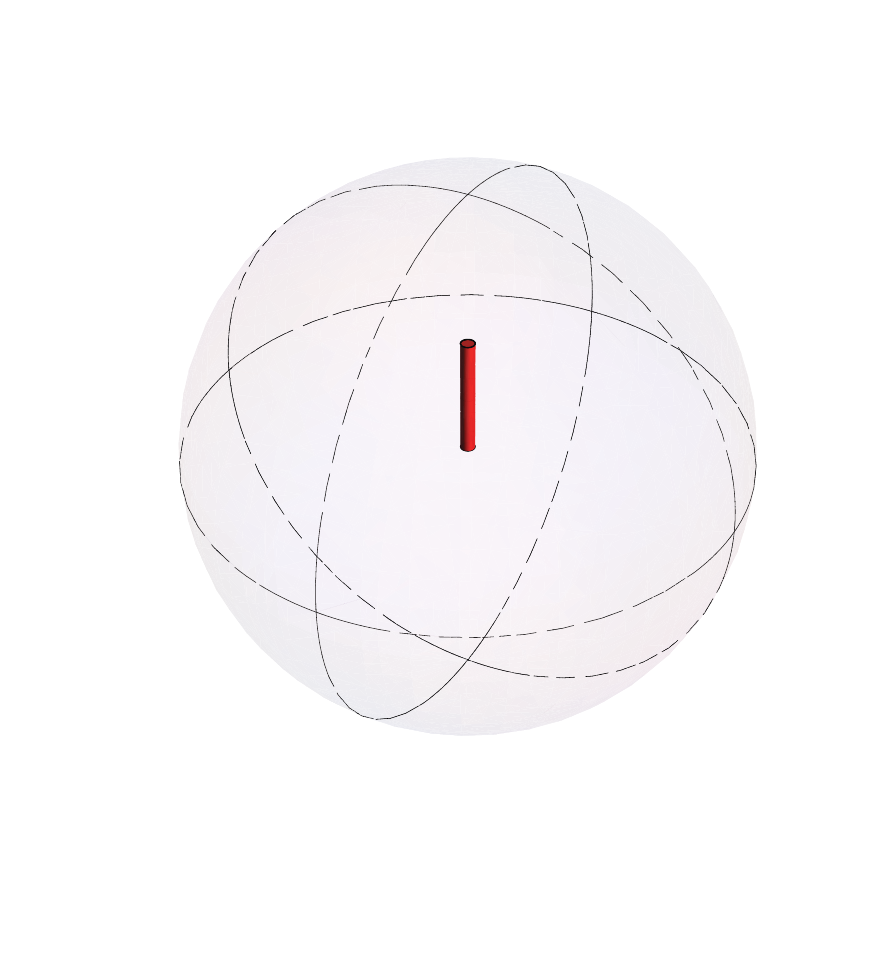}} 
	\caption[]{Evolution of the Bloch Vector
                of the reduced dynamics of qbit $1$ in the initial
                state $\rho_{1}=\ket{0}\bra{0}$ as a function of
              time.}
	\label{pic1} 
\end{figure} 
\end{center}
 
%------------------------------------------------ 
 
\subsubsection{Noiseless Quantum Simulation} 
 
Here we consider the cases where no bath is present, but 
different initial states are considered.  
Three cases are considered corresponding to 
three types of different initial states used in the simulation: 
\begin{itemize} 
\item[A.1] Pure states
\beq 
\left|\psi(0)\right\rangle=\left|00\right\rangle,\left|01\right\rangle,\left|10\right\rangle,\left|11\right\rangle. 
\eeq 
Density operator calculated as $\rho(0)=\left|\psi(0)\right\rangle\left\langle\psi(0)\right|$. 
\item[A.2] Entangled states  
\beq  
\left|\psi(0)\right\rangle=\alpha_0\left|01\right\rangle+\alpha_1\left|10\right\rangle, 
\eeq 
where $\alpha_{0}^{2}+\alpha_{1}^2 = 1$, and the density operator is given by $\rho(0)=\left|\psi(0)\right\rangle\left\langle\psi(0)\right|$. 
\item[A.3] Correlated states 
\beq 
\rho(0)=(1-p)(\rho_{1}^{I}\otimes\rho_{2}^{I})+p(\rho_{1}^{II}\otimes\rho_{2}^{II}), 
\eeq 
where $\rho_{1}^{I}$ and $\rho_{2}^{I}$ are the density operators 
corresponding to some initial state of the impurity (``spin-down''/occupied) and
fermion (``spin-up''/unoccupied), 
respectively, and $\rho_{1}^{II}$ and $\rho_{2}^{II}$ correspond to 
the other initial state of the impurity (``spin-up''/unoccupied) and
fermion (``spin-up''/unoccupied).  
\end{itemize} 
We represented the initial state of the impurity 
in terms of its $x$, $y$ and $z$ projections of the Bloch vector. The magnitude 
of each component of the projections, $a_{i}$, can be obtained by performing the 
partial trace over everything else except qbit $1$, as 
$a_{i}=\tr{[\sigma_{i}(\rho(0))]}$.  

\begin{figure}[ht!] 
  \begin{center} 
    \includegraphics[width=8cm]{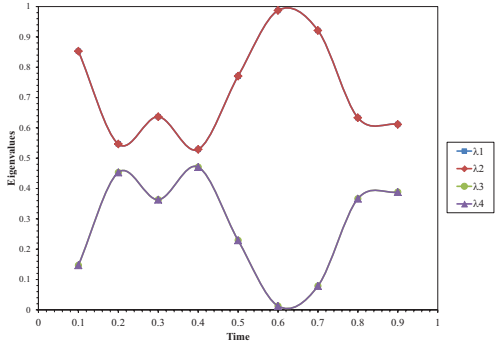} 
    \caption{Eigenvalues of the dynamical map of the reduced dynamics of qbit $1$ 
               in the initial state $\ket{\psi}=\ket{01}$ in the closed system. 
	The parameters are $\epsilon = 8$, $\varepsilon = -2$, $V = 4$, for the time interval 
          $\Delta t \in [0.1,0.9]$.} 
    \label{fig1} 
  \end{center} 
\end{figure}

First consider an initial density operator
\beq 
\rho_{S}(0)=\frac{1}{2} \left(\Bid + \vec{a_{i}}\cdot \vec{\sigma_{i}} \right) \nonumber. 
\eeq 
In this case, case A.1,  
\beq 
\rho_{S}(0)=\frac{1}{2} \left(\Bid + a_{3}\sigma_{z} \right), \nonumber 
\eeq 
where $a_3$ represents a real constant that is equal to, or less than, the radius of 
the Bloch sphere (i.e. $0 \le a_{3} \le 1$). It represents 
the projection along the $z$ axis. 
The final state was obtained through the reduced dynamics of $\rho_S$  
after the evolution: 
\beq 
\rho_{S}(t)=\tr{[\rho_{S}(0) \left(U \rho(0) U^{\dagger}\right)]}. \nonumber 
\eeq 
When the initial states $\rho_S(0)$ only had a $z$ component, the
final states $\rho_S(t)$ only had a $z$ component as well  
\beq 
\rho_{S}(t)=\frac{1}{2} \left(\Bid + b_{3} \sigma_{z} \right), \nonumber 
\eeq 
where $b_3$ is another real constant that is subject to $0 \le b_{3}
\le 1$.  The value of $b_3$ depends on $a_3$ and on the parameters 
$\epsilon$, $\varepsilon_{k_i}$, $V$ and $t$.  
When states with only a $z$ component are input, the final states also
only have a $z$ component.  This is consistent with the hopping model
where the ``spin-down'' corresponds to the state being occupied.  
The evolution is described by the dynamical map 
\bea 
\label{eq:Bmap1} 
B = \left(\begin{array}{cccc} 
                 \frac{1+b_3}{2} &        0        &       0         &       0\\ 
                 	 0       & \frac{1+b_3}{2} &       0         &       0\\ 
                         0       &        0        & \frac{1-b_3}{2} &       0\\ 
		         0       &        0        &       0         & \frac{1-b_3}{2} \end{array}\right). 
\eea 
The the eigenvalues of the map are plotted as functions of time in Figure ~(\ref{fig1}).   

The case of a maximally entangled state case A.2 also has 
a similar form with only a $z$ component.  
Therefore the map is also given by a form similar to Eq.~(\ref{eq:Bmap1}). 
As can be seen in Fig.~(\ref{fig1}), the eigenvalues of the map 
correspond to a completely positive evolution, even though the quantum
discord of the initial state was a maximum.  
Similarly, for case A.3.  We therefore note, for later
reference, that in these cases 
all states have only a $z$ component in the initial and final states of
the system.  Thus there is only this standard interpretation of the
hopping model Hamiltonian when there is no external noise.

%------------------------------------------------ 
  
\subsubsection{Simulation with noise from spin bath} 
 
In this section we present the results for systems governed by the Hamiltonians 
in Eqs.~(\ref{eq:controlh2s}) and (\ref{eq:controlh1s}). 
The goal is to simulate a two body problem, so we used the same control 
sequence in Eq.~(\ref{eq:controlseq}). However, 
the initial state of a ``bath'' of two particles was included in the total
system Hamiltonian.  As in the simulation that had no external noise,
we chose different initial configurations,  Explicitly, including the
bath qbits these are: 
\begin{itemize} 
\item[A.4] Pure states 
\beq 
\left|\psi(0)\right\rangle=\left|0011\right\rangle,\left|0111\right\rangle,\left|1011\right\rangle,\left|1111\right\rangle, 
\eeq 
and density operator $\rho(0)=\left|\psi(0)\right\rangle\left\langle\psi(0)\right|$. 
\item[A.5] Entangled states 
\beq  
\left|\psi(0)\right\rangle=\alpha_0\left|0111\right\rangle+\alpha_1\left|1011\right\rangle 
\eeq 
Where $\alpha_{0}^{2}+\alpha_{1}^2={1}$, and the density operator is given by $\rho(0)=\left|\psi(0)\right\rangle\left\langle\psi(0)\right|$ 
\item[A.6] Correlated states 
\beq 
\rho(0)=\left((1-p)(\rho_{1}^{I}\otimes\rho_{2}^{I})+p(\rho_{1}^{II}\otimes\rho_{2}^{II})\right)\otimes(\left|1\right\rangle\left\langle
  1\right|)\otimes(\left|1\right\rangle\left\langle 1\right|).  
\eeq 
\end{itemize}

The fact that the states only had a component in the $z$ direction 
and only interact with the bath via $zz$ couplings resulted in results that 
were very similar to the ones in the previous section. 
The initial state of qbit $1$ (the impurity) can again be written in Pauli notation as: 
\beq 
\rho_{S}(0)=\text{Tr}_E \rho(0)=\frac{1}{2}(\Bid + a_3 \sigma_z).
\eeq 
The final state is obtained by tracing over the bath degrees of freedom 
\beq 
\rho_1(t)=\text{Tr}_E \left(U\rho(0) U^\dagger\right)=\frac{1}{2}(\Bid + b_3 \sigma_z), 
\eeq 
$b_3$ is another real constant that can be positive or negative, depending on the direction 
of the Bloch vector of the final state along the $z$ axis. 
  
The most general dynamical map has the same form as the map in Eq.~(\ref{eq:Bmap1}), 
\bea 
B = \left(\begin{array}{cccc} 
                 \frac{1+b_3}{2} &        0        &       0         &       0\\ 
                 	 0       & \frac{1+b_3}{2} &       0         &       0\\ 
                         0       &        0        & \frac{1-b_3}{2} &       0\\ 
		         0       &        0        &       0         & \frac{1-b_3}{2} \end{array}\right). 
\eea 
 
\begin{figure}[ht!] 
 \begin{center} 
    \includegraphics[width=8cm]{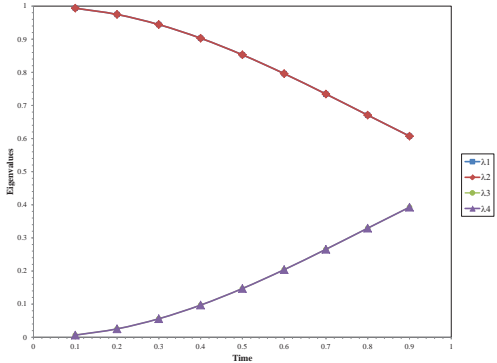} 
   \caption{Eigenvalues of the dynamical map of the reduced dynamics of qbit $1$. Two additional 
	qbits are interacting via $zz$ coupling with qbit $2$, $J_{zz}=8$. The initial state of the total system and bath is      
           $\ket{\psi}=\ket{0111}$.  
	The system parameters are $\epsilon=-8$, $\varepsilon=-2$ and $V=4$, in the time interval $t \in   
            [0.1,0.9]$.}  
   \label{fig2} 
 \end{center} 
\end{figure}

\begin{figure}[ht!] 
 \begin{center} 
    \includegraphics[width=8cm]{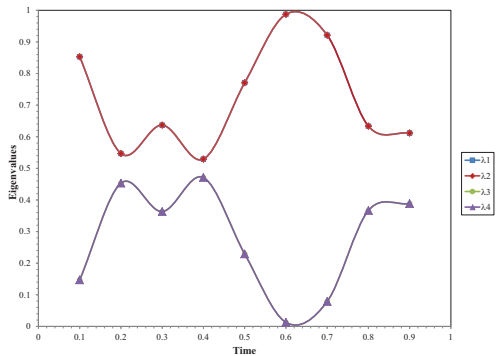} 
   \caption{Eigenvalues of the dynamical map of the reduced dynamics of qbit $1$. An additional 
	qbit is interacting via $zz$ coupling with qbits $1$ and $2$, $J_{zz}=1/10$. The initial state is 
	$\ket{\psi}=\ket{011}$. The system parameters are $\epsilon=-8$, $\varepsilon=-2$ and $V=4$,  
	for times $t \in [0.1,0.9]$.}  
   \label{fig3} 
 \end{center} 
\end{figure}

We observed that the coupling $J_{zz}$
has an effect in the rate of change of the state of qbit $1$, which is presented in the 
results for the calculation of the time correlation function.  
For purposes of comparison, the parameters of the system were the same as the  
results above. 
In Figs.~(\ref{fig2}) and~(\ref{fig3}), the eigenvalues of 
$B$ are plotted with the couplings to the particles of the spin bath  
being $J_{zz}=8$ and $J_{zz}=\frac{1}{10}$ respectively. 
 
 %%%%%%%%%%%%%%%%%%%%%%%%%%%%

Figs.~(\ref{fig1}), (\ref{fig2}) and (\ref{fig3})  
show the evolution of the same initial state but each has a different environment.  
Being states initially in the $z$ direction, the dynamics are
completely positive since the 
interaction with the bath is a $zz$ coupling.  However, it does
change the hopping rate.  
In Fig.~(\ref{fig2}) this is particularly noticeable due to the choice of the coupling. The state
of the impurity does not transfer as easily due to the strong correlations generated by
the interaction with the spin bath.  
In Fig.~(\ref{fig3}) the situation 
is different. In this case the eigenvalues remained the same
regardless of the strength of the coupling with the environment. Having a 
single extra particle interacting with both qbits with the same strength 
would mean they both have the same 
interaction with the bath.

%------------------------------------------------------------------------ 
 
\subsection{Arbitrary initial direction of the Bloch vector} 
 
Noise in the initialization of the state could result in a direction
for the Bloch vector which is not in the $z$ direction.  
States that have an $x$ or a $y$  
component to their polarization vector, or Bloch  
vector, exhibit precession and approximate more accurately 
what happens in a real experimental situation.  In this case, if some 
initial error on the system has a component in the $x$ or $y$  
direction, Larmor precession will be present.  This is often observed 
in a NMR device under general circumstances and leads to noise in the 
system.  Here we consider an initial state with a component of the Bloch  
vector in the $x$ direction.  Clearly a $y$ component is not necessary,  
and only specifies a different initial condition for the angle  
since the system will precess. 

%------------------------------------------------------------------------- 
 
\subsubsection{Noiseless Quantum Simulation} 
 
The initial states were chosen to have a component in the $x$
direction; the components in $x$ and $z$ were selected such that 
the magnitude of the Bloch vector is close to $1$ emulating a small
error in the initialization. 
Explicitly, the different initial configurations were: 
\beq 
\rho_{1}(0)=\frac{1}{2}(\Bid + a_1 \sigma_x + a_3 \sigma_z), \nonumber 
\eeq 
or  
\beq 
\rho_{2}(0)=\frac{1}{2}(\Bid + a_1 \sigma_x - a_3 \sigma_z), \nonumber 
\eeq 
where $\rho_1$ is the state of the impurity, $\rho_2$ is the state 
of the fermion and the $a_i$ are subject to  $0\le
\sqrt{a_{1}^2 + a_{3}^2} \le 1$.  Therefore, the total initial state is  
\beq 
\rho(0)=\rho_{1}(0)\otimes\rho_{2}(0) \nonumber.   
\eeq 

The final state of the impurity was, once again, obtained by doing a partial trace 
over the degrees of freedom of the fermion 
\beq 
\rho(t)=\text{Tr}_E \left(U\rho(0) U^\dagger\right)=\frac{1}{2}(\Bid + b_1 \sigma_x + b_2 \sigma_y + b_3 \sigma_z). 
\eeq 

The map $B$ is given by  
\beq 
B = \left(\begin{array}{cccc} 
\label{eq:Bmap3} 
                 \frac{1+b_3}{2} &        0        &       0         &      \frac{-ib_2}{a_1} \\ 
                 	 0       & \frac{1+b_3}{2} & \frac{b_1}{a_1}&       0              \\ 
                         0       &\frac{b_1}{a_1}& \frac{1-b_3}{2} &       0              \\ 
		 \frac{ib_2}{a_1} &        0        &       0         & \frac{1-b_3}{2} \end{array}\right). 
\eeq 
The eigenvalues of $B$ are given by 
\bea 
\lambda_1 = \frac{a_1 - \sqrt{4 b^{2}_{1}+ a^{2}_{1}b^{2}_{3}}}{2 a_1},  
%	\nonumber \\ 
\;\; &&\lambda_2 = \frac{a_1 + \sqrt{4 b^{2}_{1}+ a^{2}_{1}b^{2}_{3}}}{2 a_1},  
	\nonumber \\ 
\lambda_3 = \frac{a_1 - \sqrt{4 b^{2}_{2}+ a^{2}_{1}b^{2}_{3}}}{2 a_1} , 
%	\nonumber \\ 
\;\; &&\lambda_4 = \frac{a_1 + \sqrt{4 b^{2}_{2}+ a^{2}_{1}b^{2}_{3}}}{2 a_1},
\eea  
\vspace{.3in}
%{\onecolumngrid{
where
\begin{widetext}
\bea 
b_1 &=& \left\{\cos{\left(\frac{1}{2}t(\epsilon+\varepsilon_{k_0})\right)} 
\cos{\left(\frac{1}{2}t\sqrt{4 V^{2}+(\epsilon-\varepsilon_{k_0})^{2}} \right)} 
-\sin{\left( \frac{1}{2}t(\epsilon+\varepsilon_{k_0})\right)} 
\left[\frac{(\epsilon-\varepsilon)\sin{\left(\frac{1}{2}t\sqrt{4V^{2}+(\epsilon-\varepsilon_{k_0})^{2}}\right)}} 
{\sqrt{4V^{2}+(\epsilon-\varepsilon_{k_0})^{2}}} \right]\right\}a_1, \nonumber 
\\ 
b_2 &=& \left\{-\sin{\left(\frac{1}{2}t(\epsilon+\varepsilon_{k_0})\right)} 
\cos{\left(\frac{1}{2}t\sqrt{4 V^{2}+(\epsilon-\varepsilon_{k_0})^{2}} \right)} 
-\cos{\left( \frac{1}{2}t(\epsilon+\varepsilon_{k_0})\right)} 
\left[\frac{(\epsilon-\varepsilon)\sin{\left(\frac{1}{2}t\sqrt{4V^{2}+(\epsilon-\varepsilon_{k_0})^{2}}\right)}} 
{\sqrt{4V^{2}+(\epsilon-\varepsilon_{k_0})^2}} \right]\right\}a_1 \nonumber \\
\text{and}&& \nonumber \\  
b_3 &=& \frac{2(-1+a_3) V^2 + a_3 (\epsilon-\varepsilon)^2 + (1+a_3) V^2  
\cos{\left(\frac{1}{2}t\sqrt{4 V^2 + (\epsilon - \varepsilon)^2}\right)}}{4 V^2 + (\epsilon - \varepsilon)^2}. 
\eea 
%}}%
\end{widetext}

\begin{center}
\begin{figure}[htp] 
	\subfloat[t=0]{\includegraphics[scale=0.3]{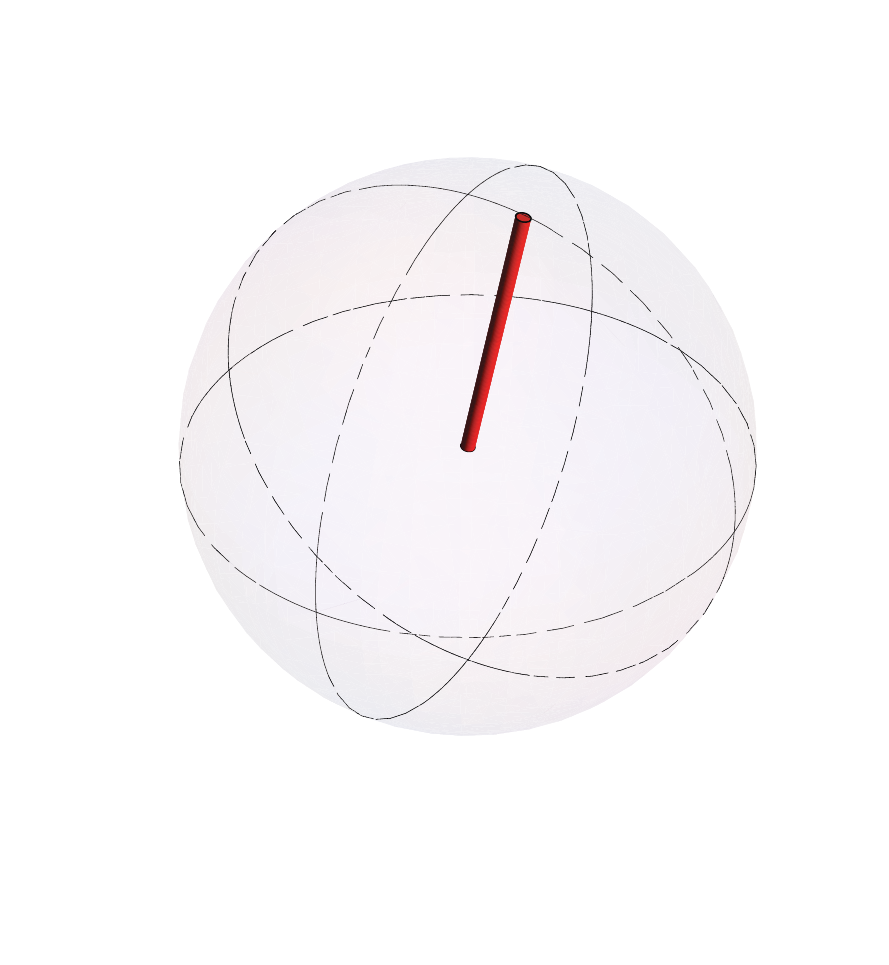}} 
	\subfloat[t=0.1]{\includegraphics[scale=0.3]{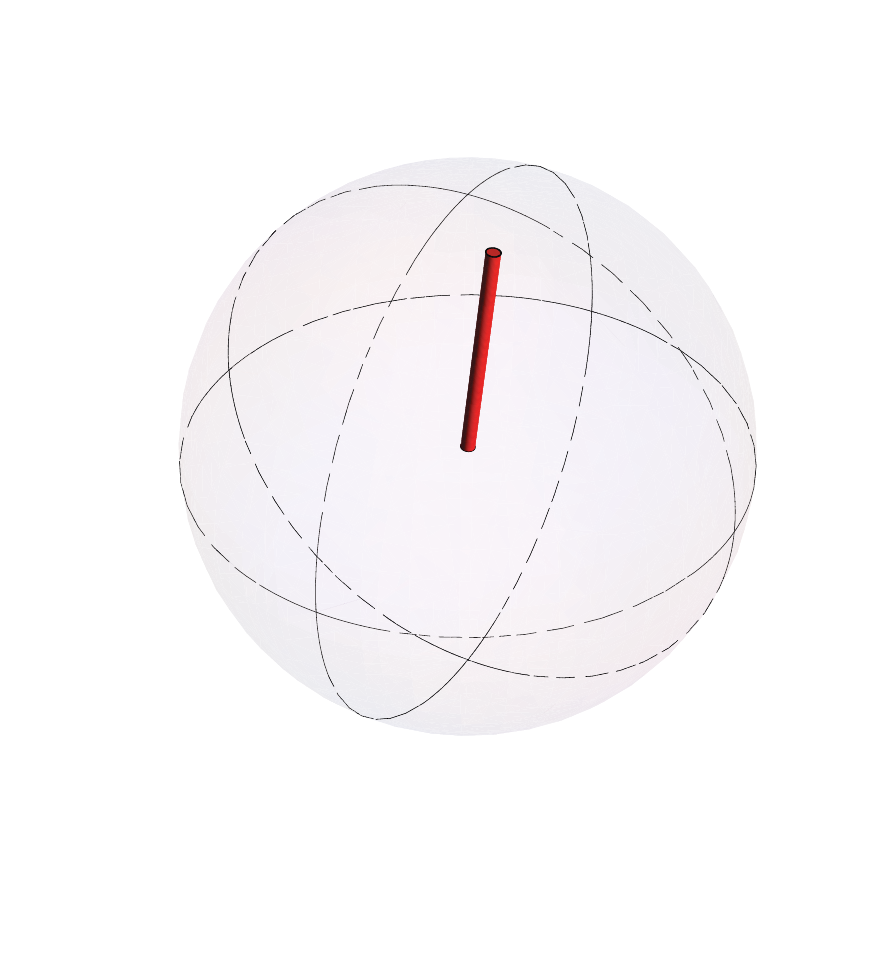}} 
	\quad
        \subfloat[t=0.2]{\includegraphics[scale=0.3]{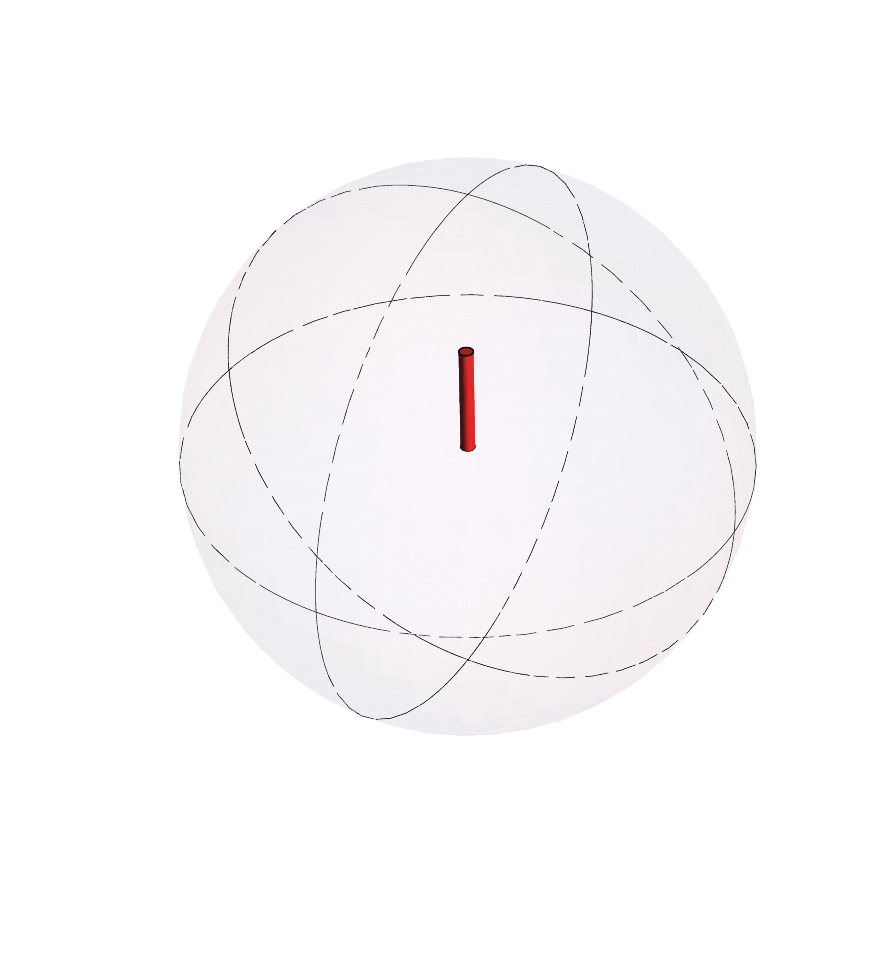}} 
	\subfloat[t=0.3]{\includegraphics[scale=0.3]{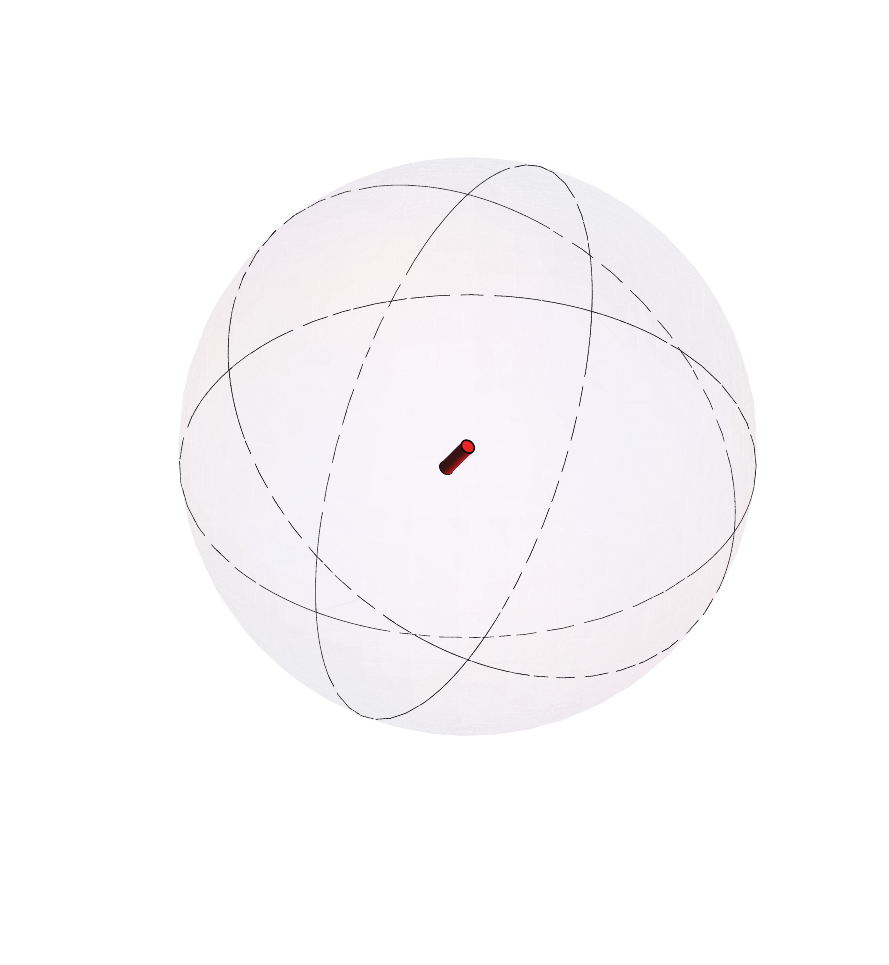}} 
        \quad
	\subfloat[t=0.4]{\includegraphics[scale=0.3]{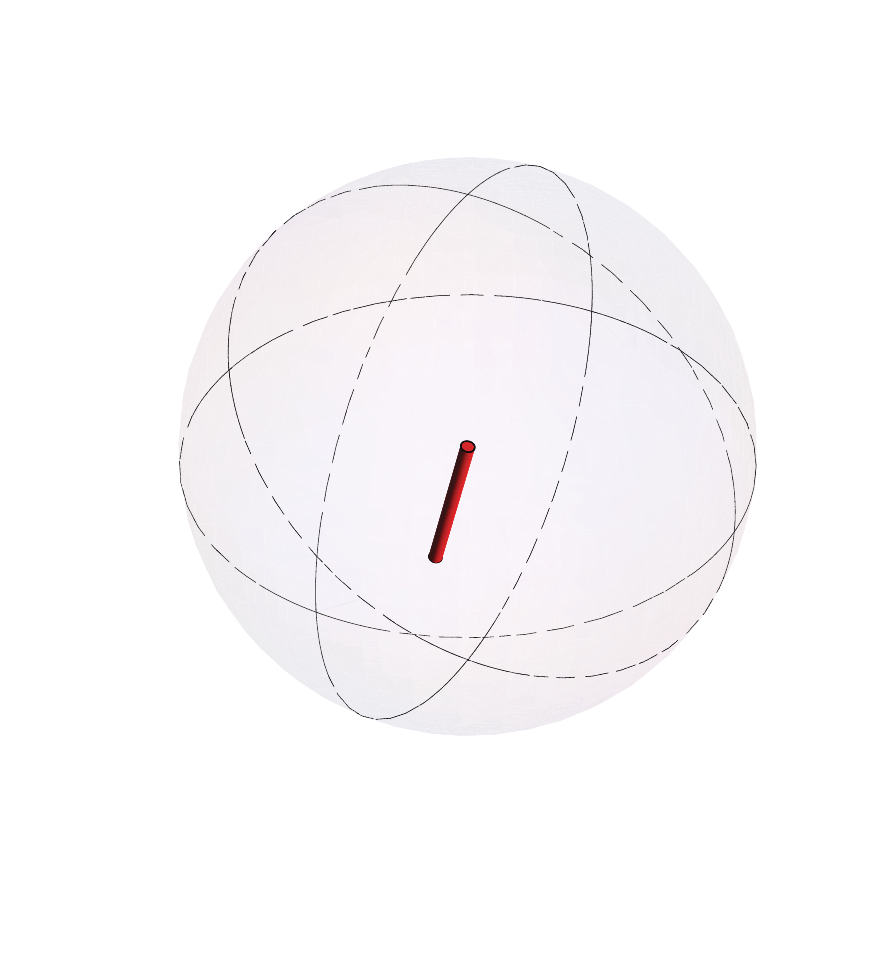}}  
	\subfloat[t=0.5]{\includegraphics[scale=0.3]{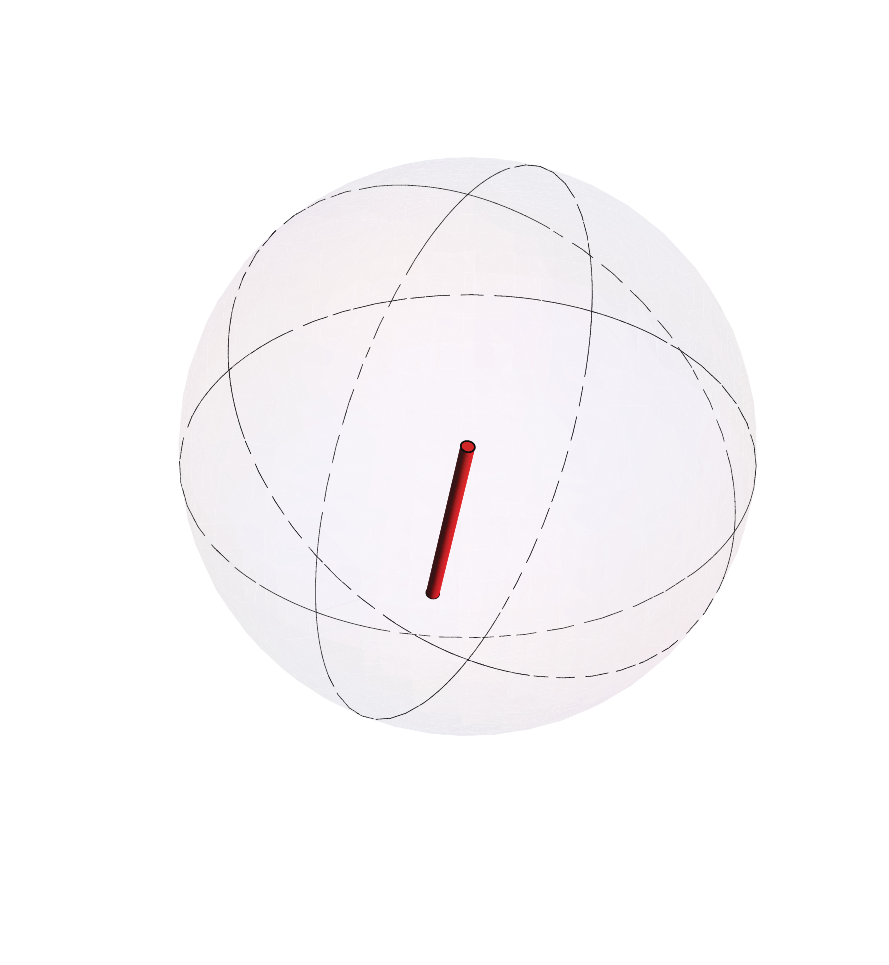}} 
	\quad
        \subfloat[t=0.6]{\includegraphics[scale=0.3]{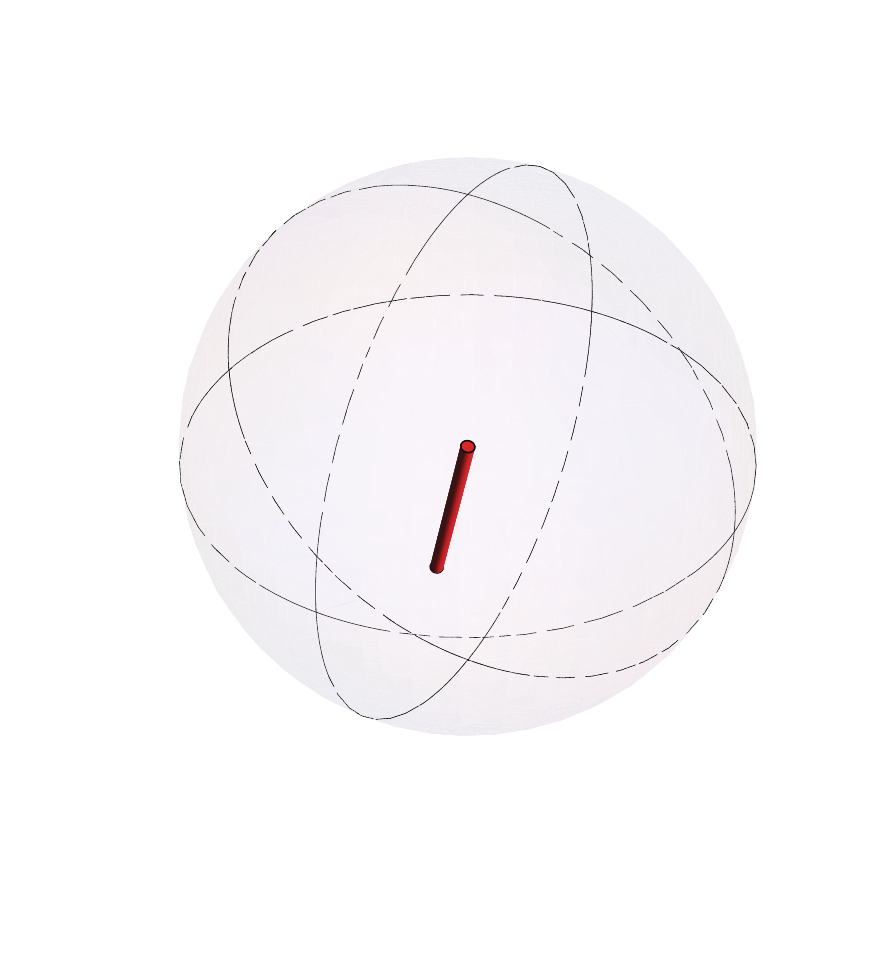}} 
	\subfloat[t=0.7]{\includegraphics[scale=0.3]{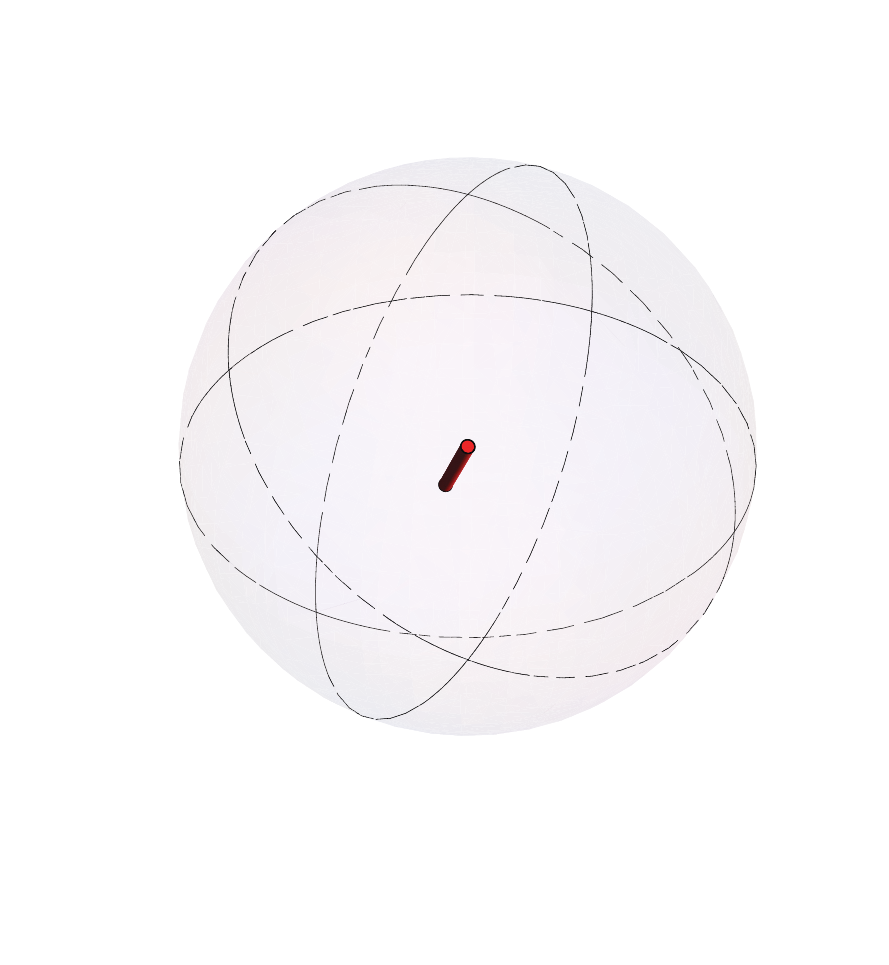}} 
%	\quad
%      \subfloat[t=0.8]{\includegraphics[scale=0.3]{xzstate9.pdf}} 
%	\subfloat[t=0.9]{\includegraphics[scale=0.3]{xzstate10.pdf}} 
	\caption[]{Animation of the evolution of the Bloch Vector
                of the reduced dynamics of qbit $1$ in the initial
                state $\rho_1=\frac{1}{2} \left(\Bid + 0.2 \sigma_x +0.97\sigma_z \right)$} 	\label{pic2} 

\end{figure} 
\end{center}

\begin{figure}[ht!] 
  \begin{center} 
    \includegraphics[width=8cm]{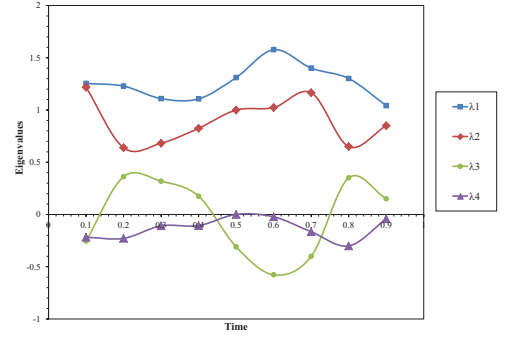} 
    \caption{Eigenvalues for the dynamical map of the reduced dynamics of qbit $1$. 
           The initial state of the system is $\rho_1=\frac{1}{2}
           \left(\Bid + 0.2 \sigma_x +0.97\sigma_z \right)$, 
           $\rho_2 = \frac{1}{2} \left(\Bid - \sigma_z \right)$. 
	The parameters of the system are $\epsilon = -8$, $\varepsilon = -2$, $V = 4$. Evaluated in the time 
            interval $t \in [0.1,0.9]$.} 
    \label{fig4} 
  \end{center} 
\end{figure}

Note that if $a_1 \mapsto 0$, then $b_1$ and $b_2$ are $0$. The factor 
$a_1$ in the denominator of the eigenvalues is eliminated using  
l'Hospital's rule, and that yields  
\bea 
\lambda_1 = \frac{1 - b_3}{2}, \;\; 
%	\nonumber \\ 
&&\lambda_2 = \frac{1 + b_3}{2},  
	\nonumber \\ 
\lambda_3 = \frac{1 - b_3}{2}, \;\;
%	\nonumber \\ 
&&\lambda_4 = \frac{1 + b_3}{2},  
	\nonumber \\ 
\eea 
which are the same as the eigenvalues of the map in Eq.~(\ref{eq:Bmap1}). 
The eigenvalues of $B$ when $a_{1}>0$ are shown in Figure~(\ref{fig4}). 
%%%%%%%%%%%%%%%%%%%%%%%%%%%%%%%%%%%%%%%%%%%%%%%%%%%%%%%%%%%%%%%%%%%%%%
In Fig.~(\ref{fig4}), the dynamics of the system are positive but not completely 
positive. This system is not in contact with a bath or reservoir, but it consists of two 
particles. This is a case of initial correlations between particles in 
the system, which are errors for this model since 
correlations should not be present in initial state preparation.
The general observation was that when the initial state has a component  
of the Bloch vector in $x$ or $y$ as well as one in $z$, the result is
a non completely positive map.

%------------------------------------------------------------------------------------------------------------ 
 
\subsubsection{Simulation with noise from spin bath} 
 
\begin{figure}[ht!] 
 \begin{center} 
    \includegraphics[width=8cm]{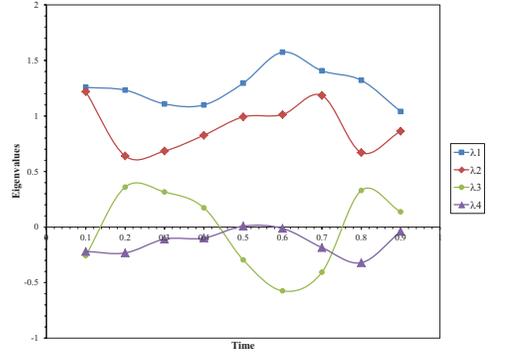} 
   \caption{Eigenvalues of the dynamical map for the reduced dynamics
     of qbit $1$. The initial state of the system is 
     $\rho_1=\frac{1}{2} \left(\Bid + 0.2 \sigma_x +0.97\sigma_z
     \right)$, $\rho_2 = \frac{1}{2} \left(\Bid - \sigma_z
     \right)$, $\rho_3 = \frac{1}{2} \left(\Bid - \sigma_z
     \right)$, $\rho_4 = \frac{1}{2} \left(\Bid - \sigma_z
     \right)$.  The system parameters are 
   $\epsilon = -8$, $\varepsilon = -2$, $V = 4$, $J_{zz} =
   \frac{1}{10}$ evaluated in the time interval $t \in [0.1,0.9]$.}  
   \label{fig5} 
 \end{center} 
\end{figure}

\begin{figure}[ht!] 
 \begin{center} 
    \includegraphics[width=8cm]{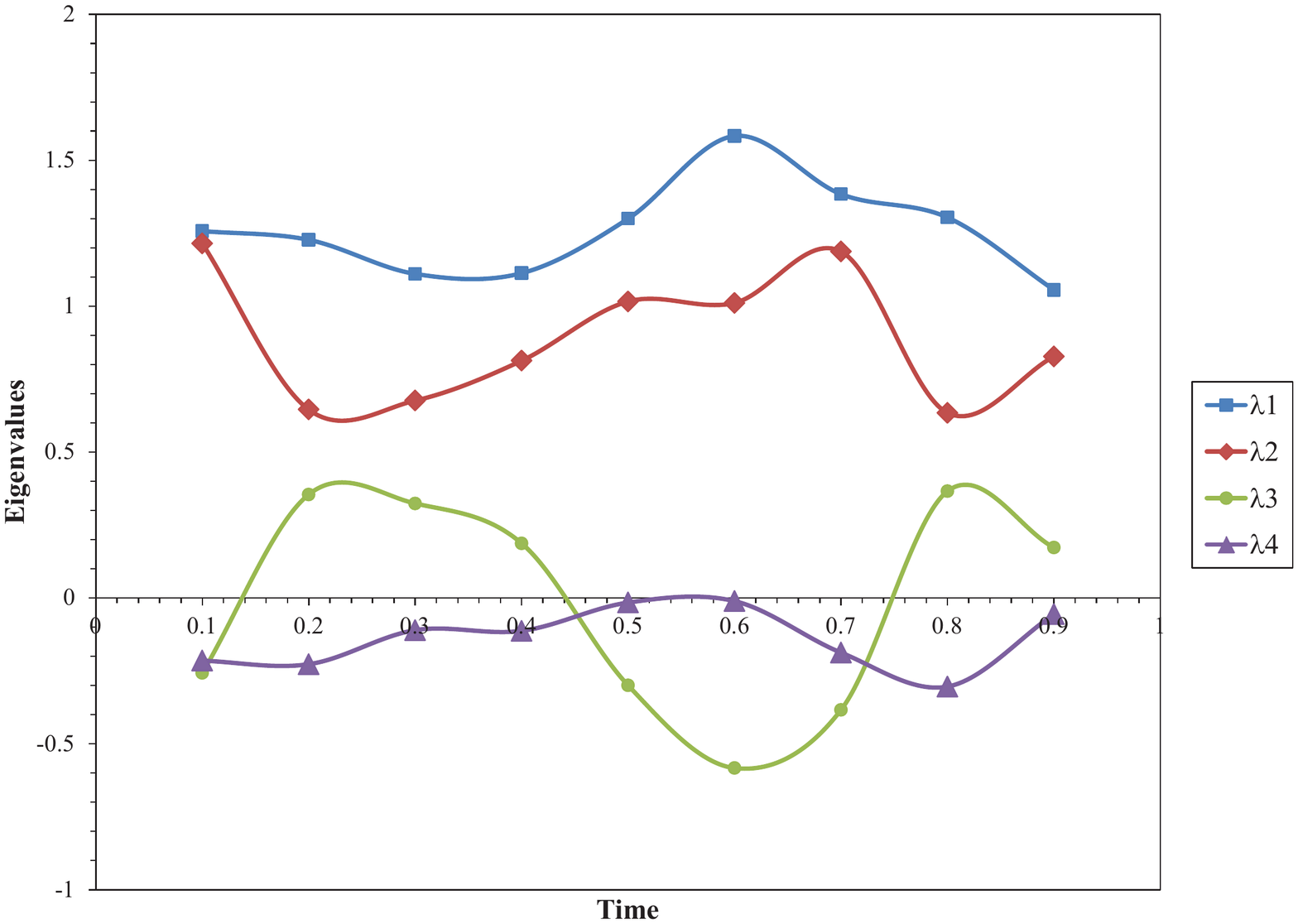} 
   \caption{Eigenvalues of the dynamical map for the reduced dynamics
     of qbit $1$.  The initial state of the system is
     $\rho_1=\frac{1}{2} \left(\Bid + 0.2 \sigma_x +0.97\sigma_z
     \right)$, $\rho_2 = \frac{1}{2} \left(\Bid - \sigma_z
     \right),$   $\rho_3 = \frac{1}{2} \left(\Bid - \sigma_z
     \right)$. The system parameters are 
   $\epsilon = -8$, $\varepsilon = -2$, $V = 4$, $J_{zz} =
   \frac{1}{10}$ evaluated in the time interval $t \in [0.1,0.9]$.}  
   \label{fig6} 
 \end{center} 
\end{figure}

The results in this subsection are generated from adding the qbits in
the spin bath, and using the following initial states  
\beq 
\rho(0)=\rho_{1}(0)\otimes\rho_{2}(0)\otimes(\left|1\right\rangle\left\langle 1\right|)\otimes(\left|1\right\rangle\left\langle 1\right|), 
\eeq 
where 
\beq 
\rho_{1}(0)=\frac{1}{2}(\Bid + a_1 \sigma_x + a_3 \sigma_z), 
\eeq 
and 
\beq 
\rho_{2}(0)=\frac{1}{2}(\Bid + a_1 \sigma_x - a_3 \sigma_z). 
\eeq 
The reduced dynamics of $S$ are given by 
\beq 
\rho(t)=\text{Tr}_E \left(U\rho(0) U^\dagger\right)=\frac{1}{2}(\Bid + b_1 \sigma_x + b_2 \sigma_y + b_3 \sigma_z) ,
\eeq 
with a $B$ map of the same for as that in Eq.~(\ref{eq:Bmap3}),  
\bea 
B = \left(\begin{array}{cccc} 
                 \frac{1+b_3}{2} &        0        &       0         &      \frac{-ib_2}{a_1} \\ 
                 	 0       & \frac{1+b_3}{2} & \frac{b_1}{a_1}&       0              \\ 
                         0       &\frac{b_1}{a_1}& \frac{1-b_3}{2} &       0              \\ 
		 \frac{ib_2}{a_1} &        0        &       0 
                 & \frac{1-b_3}{2} \end{array}\right).  
\eea 
Once again, the noise, which has the form of purely $zz$ couplings, 
caused variations in the parameters, mostly in the rate of change 
of the state of qbit $1$. 
The eigenvalues for a system with two spins 
interacting with the fermion only and for one spin interacting with both 
particles in the system are presented in Figs.~(\ref{fig5}) and~(\ref{fig6}). 
 
%%%%%%%%%%%%%%%%%%%%%%%%%%%%%%%%%%%%%%%%%%%%%%%%%%%%%%%%%%%%%%%%%%%%%%%
In Figs.~(\ref{fig5}) and (\ref{fig6}) the reduced dynamics are not completely  
positive.  This is due to the initial state of the impurity
site (qbit $1$) having a component  
of its Bloch vector in the $x$ direction. The  
algorithm was designed to have an initial state where one
of the two state systems is in the up state 
and the rest are in the down state. Dynamical maps obtained through quantum process 
tomography can present discrepancies if the initial states are prepared through different 
experimental methods \cite{Kuah:07}.  Thus the $x$ component represents a
preparation error which gives rise to a non-completely positive map
like in the previous case.  

%----------------------------------------------------------------------- 
 
\subsection{Time correlation function} 
 
\begin{figure}[ht!] 
 \begin{center} 
    \includegraphics[width=8cm]{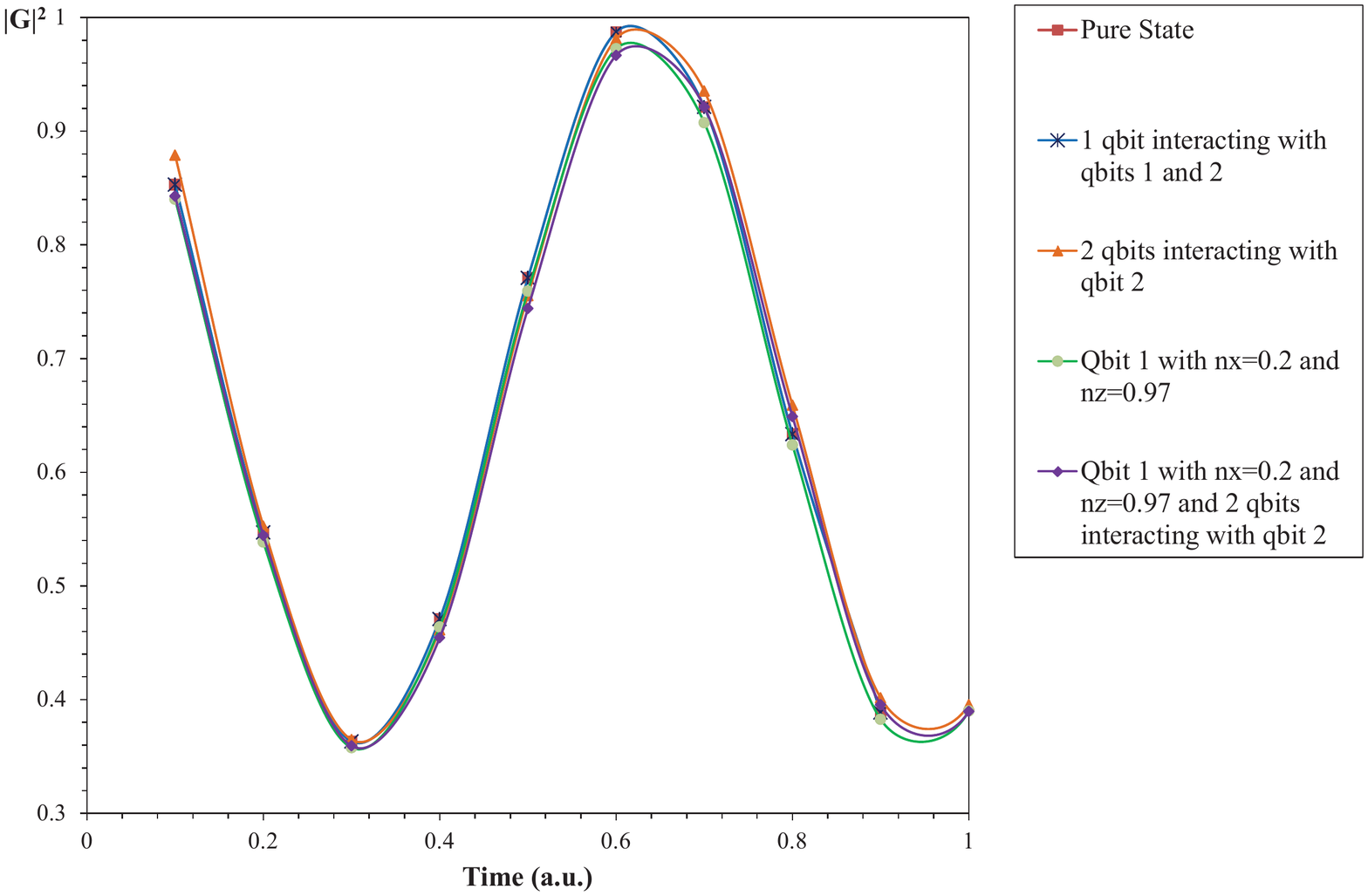} 
   \caption{Time correlation function of the reduced dynamics of qbit $1$. The system parameters are 
   $\epsilon = -8$, $\varepsilon = -2$, $V = 4$ and time interval $t \in [0.1,0.9]$. 
   The results represent the closed system, the system where qbit $2$ interacts with 
   two additional qbits, the system in which an additional qbit that interacts with qbits $1$ and $2$.
   This was done when qbit $1$ was in the initial states $\rho=\ket{0}\bra{0}$ and $\rho=\frac{1}{2}\left(\Bid+ 0.2 \sigma_{x}
   +0.97\sigma_{z}\right)$, as indicated above.}  
   \label{corr1} 
 \end{center} 
\end{figure}

\begin{figure}[ht!] 
 \begin{center} 
    \includegraphics[width=8cm]{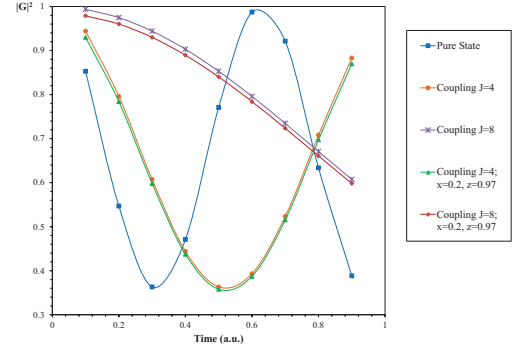} 
   \caption{Time correlation function of the reduced dynamics of qbit $1$. The system parameters 
   are $\epsilon = -8$, $\varepsilon = -2$, $V = 4$ in the time interval $t \in [0.1,0.9]$. 
   The results correspond to the closed system and the system that interacts with 
   two additional qbits, coupled only to qbit $2$. The initial state ofqbit $1$ is
   $\rho=\ket{0}\bra{0}$ for one set of results, and $\rho=\frac{1}{2}\left(\Bid+ 0.2 \sigma_{x}
   +0.97\sigma_{z}\right)$ for the other.}  
   \label{corr2} 
 \end{center} 
\end{figure}

\begin{figure}[th!] 
 \begin{center} 
    \includegraphics[width=8cm]{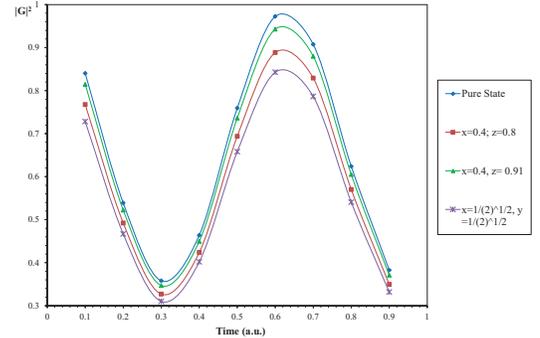} 
   \caption{Time correlation function of the reduced dynamics of qbit $1$. The system parameters
   are $\epsilon = -8$, $\varepsilon = -2$, $V = 4$ evaluated in the time interval $t \in [0.1,0.9]$. 
   The result represents the time correlation function of the closed system compared to the correlation
   function of the reduced dynamics of qbit $1$ in the initial state $\rho=\frac{1}{2}\left(\Bid+ a_{1} \sigma_{x}
   +a_{3}\sigma_{z}\right)$ for different values of $a_{1}$ and $a_{3}$.}  
   \label{corr3} 
 \end{center} 
\end{figure}

Ortiz et al.~calculated the time correlation function $C(t) = b(t)b(0)^{\dagger}$,  
and plotted the result as 
${\abs{G}}^{2}=\tr\left(\rho(t)\rho(0)\right)$ as a function of time.  
Since we want to calculate the effects of noise and  
different initial state, we followed the same procedure for the different situations. 
The results are summarized in graphs, Figs.~(\ref{corr1}), ~(\ref{corr2}) and 
~(\ref{corr3}).  
In Fig.~(\ref{corr1}), there is a slight difference between the results of the original 
system compared to those under which errors could arise due to noise and 
unknown initial states. The coupling to the environment affects how fast or slow 
qbit 1 evolves.  However, if the coupling to the bath 
is weak, these errors are not as prominent. 
There was one case in which there was no effect on the speed of change by spins in the bath.
%The system where one extra qbit interacts with qbits 1 and 2. 
%This is because the algorithm in \cite{Ortiz:01} 
%is reduced to a two-body problem because the ring problem is independent of the fermions 
%surrounding the impurity. 
 
When the initial state had a component in $x$ the resulting 
correlation functions were very close to the original 
problem. This is important because a small error like this one 
may not be easily identified in the time correlation function. 
In Fig.~(\ref{corr2}) we show how the coupling to a spin bath 
can affect the rate of change of the evolution. As mentioned  
before, these results only include $zz$ couplings. 
The strength of the couplings were adjusted in order 
to see the effects.  
 
Because quantum simulations are performed on quantum systems, where access to  
complete information about the state at all times is not available, correlations with the bath  
can be by detected by differences in the rate of change of the evolution. However, it is 
useful to also study the case in which the coupling is not only in $z$. 
In Fig.~(\ref{corr3}), we increased $a_1$, the component of the Bloch vector in $x$, 
to see how it affects the final result.  When the $x$ component of the Bloch 
vector is increased, we can see shifts in the time correlation function. 
The greater $a_1$ is, the larger the observed shift.  
This could be useful for detecting possible errors in state preparation. 
 
%----------------------------------------------------------------------- 
 
\section{Conclusions} 
 
\label{sec:concl} 
 
Interactions of quantum systems with a surrounding environment are 
undesirable for reliable quantum simulations and for quantum information 
processing in general. 
Understanding and controlling or suppressing the noise from the
environment is one of the most important objectives of studying 
open system quantum dynamics. 
Lloyd's suggestion to {\sl use} the noise to simulate the interaction 
of the system with the environment is clearly useful only in special cases.  
For some analog simulators, isolation has nearly been achieved \cite{Barreiro:01}, 
but that will not be the case for many devices. For the cases where noise suppression 
is required, understanding the noise will be necessary for implementing the 
appropriate noise control method. 

It is known that interactions with the environment can lead to correlations 
that can result in non completely positive maps.
We found that such non completely positive maps are not rare in our study of 
a very simple model of a quantum system of fermions which can readily be 
simulated on a quantum computing device, or a dedicated quantum simulator. 
This Fano-Anderson model exhibits maps which are not completely positive for 
a variety of initial states, some of which were entangled and some with 
other non-trivial quantum correlations in the sense of non-zero
quantum discord.  They were shown to arise for even a 
fairly small transversal component to an initial density matrix which is supposed 
to have its Bloch vector aligned along the $z$ axis. Thus fairly small experimental 
errors can lead to maps which are not completely positive in a rather
simple experiment. These noises also cause relatively small errors in
the final outcome of the measurement.  

Initially correlated states, if they are not so identified, 
but are instead identified improperly as arising from 
completely positive maps, may encourage an experimenter to try to
employ dynamical decoupling controls to eliminate errors.  These
controls will be ineffective in these cases. 

We have used a very specific and simple model to illustrate the effects 
of noise on the system including the presences of maps which are not 
completely positive.  However, it is important to emphasize that these 
effects are quite general and will be present in some form in many other 
quantum systems including a wide class of quantum simulations.

%----------------------------------------------------------------------- 
 
\section*{Acknowledgements} 

This material is based upon work supported by the National Science 
Foundation under Grant No. 0545798.

%------------------------------------------------------------------------ 
%------------------------------------------------------------------------ 
 %merlin.mbs apsrev4-1.bst 2010-07-25 4.21a (PWD, AO, DPC) hacked
%Control: key (0)
%Control: author (8) initials jnrlst
%Control: editor formatted (1) identically to author
%Control: production of article title (-1) disabled
%Control: page (0) single
%Control: year (1) truncated
%Control: production of eprint (0) enabled
%

%------------------------------------------------------------------------ 
 
\end{document}